\documentclass[12pt,a4paper]{article}
\pdfoutput=1

\usepackage{geometry}
\usepackage{amsmath}
\usepackage{amssymb}
\usepackage[dvips]{graphicx}
\usepackage{cite}
\usepackage{pstricks}
\usepackage{bm}
\usepackage{pbox}
\usepackage{placeins}
\usepackage{graphicx}
\usepackage{caption}
\usepackage{subcaption}
\usepackage[T1]{fontenc}
\usepackage{footnote}
\usepackage{pdfpages}
\usepackage{hhline}
\usepackage{multirow}
\usepackage{enumitem}
\usepackage[toc,page]{appendix}
\allowdisplaybreaks

\numberwithin{equation}{section}

\newcommand{\be}{\begin{equation}}
\newcommand{\ee}{\end{equation}}
\newcommand{\bea}{\begin{eqnarray}}
\newcommand{\eea}{\end{eqnarray}}

\newcommand{\beq}{\begin{equation}}
\newcommand{\eeq}{\end{equation}}
\newcommand{\bet}{\begin{itemize}}
\newcommand{\eet}{\end{itemize}}
\newcommand{\ben}{\begin{enumerate}}
\newcommand{\een}{\end{enumerate}}
\newcommand{\bem}{\begin{pmatrix}}
\newcommand{\eem}{\end{pmatrix}}

\newcommand{\non}{\nonumber}

\definecolor{MyDarkBlue}{rgb}{0.1, 0.1, 0.8}
\definecolor{MyLightBlue}{rgb}{0.22,0.51,0.9}
\usepackage[colorlinks=true,linkcolor=black,citecolor=MyDarkBlue,
urlcolor=MyLightBlue,bookmarksnumbered=true,bookmarksopen]{hyperref}
\hypersetup{colorlinks, citecolor=red,linkcolor=MyDarkBlue, urlcolor=MyLightBlue}

\newcommand\scalemath[2]{\scalebox{#1}{\mbox{\ensuremath{\displaystyle #2}}}}

\begin{document}

\numberwithin{equation}{section}

\vspace*{-0.2in}
\begin{flushright}
OSU-HEP-18-08
\end{flushright}
\vspace{0.5cm}
\begin{center}

{\Large\bf Peccei-Quinn Symmetry and Nucleon Decay\\[0.1in] in Renormalizable SUSY \boldmath{$SO(10)$}}
\vspace{1cm}

\renewcommand{\thefootnote}{\fnsymbol{footnote}}
\centerline{
{}~{\bf K.S. Babu}$^{a,}$\footnote{E-mail: \textcolor{MyLightBlue}{babu@okstate.edu}},
{}~{\bf Takeshi Fukuyama}$^{b,}$\footnote{E-mail: \textcolor{MyLightBlue}{	fukuyama@se.ritsumei.ac.jp}},
{}~{\bf Saki Khan}$^{a,}$\footnote{E-mail: \textcolor{MyLightBlue}{saki.khan@okstate.edu}} and
{}~{\bf Shaikh Saad}$^{a,}$\footnote{E-mail: \textcolor{MyLightBlue}{shaikh.saad@okstate.edu}}
}

\vspace{0.5cm}
\centerline{$^{a}${\it\small Department of Physics, Oklahoma State University, Stillwater, OK, 74078, USA }}
\centerline{$^{b}${\it\small Research Center for Nuclear Physics (RCNP), Osaka University, Ibaraki, Osaka, 567-0047, Japan}}
 \end{center}

\renewcommand{\thefootnote}{\arabic{footnote}}

\bigskip

\begin{abstract}
{\footnotesize
We suggest simple ways of implementing Peccei-Quinn (PQ) symmetry to solve the strong CP problem in renormalizable SUSY $SO(10)$ models with a minimal Yukawa sector.  Realistic fermion mass generation requires that a second pair of Higgs doublets survive down to the PQ scale.  We show how unification of gauge couplings can be achieved in this context.  Higgsino mediated proton decay rate is strongly suppressed by a factor of $(M_{\rm PQ}/M_{\rm GUT})^2$, which enables all SUSY particles to have masses of order TeV. With TeV scale SUSY spectrum, $p \rightarrow \overline{\nu} K^+$ decay rate is expected to be in the observable range.  Lepton flavor violating processes $\mu \rightarrow e\gamma$ decay and $\mu-e$ conversion in nuclei, induced by the Dirac neutrino Yukawa couplings, are found to be within reach of forthcoming experiments.
}
 \end{abstract}

\clearpage
{
\hypersetup{linkcolor=black}
\tableofcontents
}
\newpage
\setcounter{footnote}{0}

\section{Introduction}

Grand unified theories (GUTs) \cite{Pati:1974yy,Georgi:1974sy,Georgi:1974yf} are some of the best motivated extensions of the Standard Model, and have been extensively studied in the literature. Not only do they unify the various forces of nature, they also unify quarks with leptons and particles with antiparticles. Unification of gauge couplings has been known to work very well in the context of TeV scale supersymmetry (SUSY), motivated independently from the Higgs mass hierarchy perspective.  SUSY GUTs based on $SO(10)$ gauge symmetry \cite{so101,so102} would unify all members of a family of quarks and leptons, including the right-handed neutrino, into a single {\bf 16}-dimensional multiplet.  Owing to such a grouping, SUSY $SO(10)$ models are capable of explaining various features of the observed fermion mass spectrum.  In particular, renormalizable SUSY $SO(10)$ models which utilize a single ${\bf 10}$ and a single ${\bf \overline{126}}$ of Higgs fields to generate fermion masses provide an excellent fit to all of quark and lepton masses and mixings, including neutrino oscillation data, with a relatively small number of parameters \cite{Babu:1992ia,Bajc:2002iw,Fukuyama:2002ch,Goh:2003sy,
Goh:2003hf,Bertolini:2004eq,Babu:2005ia,Bertolini:2006pe,
Joshipura:2011nn,Altarelli:2013aqa,Dueck:2013gca,
Bajc:2008dc,Babu:2018tfi,Deppisch:2018flu,Fukuyama:2015kra,Fukuyama:2016vgi}.  
The value of the reactor neutrino mixing angle was predicted in these models before it was measured by the DayaBay collaboration \cite{DayaBay}, which turned out to be consistent with the prediction.  These models may be referred to as SUSY $SO(10)$ models with a minimal Yukawa sector.

GUTs do not shed much insight to the strong CP problem -- why the QCD parameter $\overline{\theta}$ takes a value less than $10^{-10}$.  Perhaps the most compelling explanation of the strong CP problem is in terms of the Peccei-Quinn (PQ) symmetry, which is a global $U(1)$ symmetry broken spontaneously by a Higgs field, and also explicitly by the QCD anomaly \cite{Peccei:1977hh}.  The breaking of the $U(1)_{\rm PQ}$ leads to a near massless scalar, the axion \cite{Weinberg:1977ma,Wilczek:1977pj,Kim:1979if,Shifman:1979if,Zhitnitsky:1980tq,Dine:1981rt}, which may constitute a fraction or the entire dark matter in the universe.  The spontaneous PQ symmetry breaking scale 
should be of order $10^{11}-10^{12}$ GeV, to be consistent with direct experimental limits as well as indirect limits from astrophysics and cosmology \cite{Kim:1986ax}.  It would be of great interest to combine PQ symmetry with SUSY $SO(10)$ models with the minimal Yukawa sector, which has not been done to date.\footnote{For recent works on non-SUSY $SO(10) \times U(1)_{\rm PQ}$ models, see Ref. \cite{bk,ringwald}.}  We undertake this task in this paper.

SUSY $SU(5)$ GUT has been extended to include the $U(1)_{\rm PQ}$ symmetry \cite{Hisano:1992ne}.  To achieve this goal in SUSY $SO(10)$ with a minimal Yukawa sector,  we identify two key ingredients: (i) an additional {\bf 10} of Higgs field, and (ii) a singlet sector that breaks the PQ symmetry in the SUSY limit.  Without the additional {\bf 10} Higgs field the color triplet partners of the Higgs bosons would survive down to the PQ scale, mediating relatively rapid proton decay (assuming TeV scale masses for SUSY particles).  While the $SO(10)$ multiplets ${\bf 126}$ and ${\bf \overline{126}}$ can be utilized to break the PQ symmetry, these fields with their PQ charges being opposite (so that they can have a mass term) would leave a linear combination of $U(1)_X$ and $U(1)_{\rm PQ}$ unbroken, where $U(1)_X$ is part of $SO(10)$ gauge symmetry.  This surviving global $U(1)$ will only be broken spontaneously at the electroweak scale, leading to a weak scale axion model \cite{Weinberg:1977ma,Wilczek:1977pj}, which is excluded by direct experiments such as $K_L \rightarrow \pi a$ searches \cite{Kim:1986ax}.

One feature that results in the PQ extension of SUSY $SO(10)$ is that an extra pair of Higgs doublets $(H_u',\,H_d')$ survives down to the PQ scale.  If their masses were at the GUT scale, the masses of either the up-type quarks or the down-type quarks (and charged leptons) would be suppressed by a factor $(M_{\rm PQ}/M_{\rm GUT})$. This would lead to unacceptably small values of $b$-quark and $\tau$-lepton masses. With 
$(H_u',\,H_d')$ having masses of order the PQ scale, $m_b$ and $m_\tau$ would not be suppressed and can be fit to their observed values.
With an intermediate scale mass for $(H_u',\,H_d')$, the unification of gauge couplings that works well in the MSSM would however be spoiled.  We show that lowering the masses of certain colored multiplets slightly from their GUT scale values (by a factor of 10 or so) can compensate for the effects of the extra doublet pair at the PQ scale.   These correlated features are also present in the SUSY $SU(5) \times U(1)_{\rm PQ}$ models \cite{Hisano:1992ne}. 

One important consequence of the PQ embedding of SUSY $SO(10)$ models is that Higgsino-mediated $d=5$ proton decay operators \cite{Weinberg:1981wj,Sakai:1981pk} become suppressed compared to the corresponding non-PQ models.  This is a great bonus, as it has been shown that these $d=5$ proton decay operators lead to rather fast proton decay in SUSY $SO(10)$ with the minimal Yukawa sector, assuming that all the SUSY particles have masses of order TeV \cite{Babu:2018tfi}.  This problem prompted the suggestion of a mini-split SUSY spectrum in Ref. \cite{Babu:2018tfi} with the gauginos having masses of order TeV and scalars having masses of order 100 TeV.  The decay rate of the proton would be suppressed by a factor of $(M_{\rm PQ}/M_{\rm GUT})^2$ compared to the results of Ref. \cite{Babu:2018tfi} in the SUSY $SO(10)$ with PQ symmetry that we present here.  This suppression would enable all SUSY particles to have masses of order TeV. This statement would be quantified later on in this paper. Within this setup we  find that the decay rate for $p \rightarrow \overline{\nu} K^+$ is within reach of ongoing and proposed experiments.  With TeV scalars, we also find that lepton flavor violating (LFV) decays $\mu \rightarrow e \gamma$ and $\mu-e$ conversion in nuclei lie in the range that may be observed in forthcoming experiments.  Such flavor violations have their origin in the  neutrino Dirac Yukawa couplings which are active between $M_{\rm GUT}$ and the $B-L$ symmetry breaking scale $v_R \sim 10^{12}$ GeV.  Renormalization group flow of SUSY parameters in the momentum range $v_R \leq \mu \leq M_{\rm GUT}$ where $\nu_R$'s are active would transfer LFV information to the sleptons, which have masses of order TeV.  This in turn would lead to LFV processes such as $\mu \rightarrow e \gamma$. The Dirac neutrino Yukawa couplings and the $B-L$ breaking scale are fixed in these models owing to the minimality of the Yukawa sector, leading to crisp predictions for LFV, which depend only on the SUSY particle masses. In this framework, the dark matter candidate can be a mixture of a neutralino-like (wino-like or Higgisno-like) WIMP and a SUSY DFSZ axion \cite{Bae:2015rra, Bae:2017hlp}. An alternative possibly is to have axino as the dark matter \cite{Choi:2013lwa}.

\section{SUSY \texorpdfstring{\boldmath{$SO(10)$}}{TEXT}  with  \texorpdfstring{\boldmath{$U(1)_{\rm PQ}$}}{TEXT}}
In this section we present a viable model that combines a global $U(1)_{\rm PQ}$ symmetry with $SO(10)$ gauge symmetry in the supersymmetric context.  The PQ symmetry solves the strong CP problem; it also enables us to realize TeV scale super-particles consistent with proton decay limits.  The model we present is an extension of the renormalizable SUSY $SO(10)$ which preserves the minimality of the Yukawa sector.  Fermion families, which belong to the {\bf 16}-dimensional representations, have Yukawa couplings in these models with a single {\bf 10} and a single ${\bf \overline{126}}$ of Higgs superfiels. The Yukawa superpotential of these models is given by:
\begin{align}\label{Yuk}
W_{\rm Yuk}&= Y^{ij}_{10}\Psi_iH\Psi_j+Y^{ij}_{126}\Psi_i\overline{\Delta}\Psi_j .
\end{align}
Here $\Psi_i$ stand for the three families of fermions in the ${\bf 16}$, $H$ is the {\bf 10}-plet of Higgs and $\overline{\Delta}$ is the ${\bf \overline{126}}$-plet of Higgs.  $Y_{10}$ and $Y_{126}$ are complex symmetric Yukawa coupling matrices, of which one can be chosen diagonal without loss of generality.  It was shown in Ref. \cite{Babu:1992ia} that this Yukawa sector can generate fermion masses consistently, since the $\overline{\Delta}$ field acquires 
a large vacuum expectation value (VEV) $v_R$ along its SM singlet direction, thus breaking $B-L$ gauge symmetry and supplying Majorana masses to the right-handed neutrinos, as well as  electroweak scale VEVs along its $SU(2)_L$ doublet directions, contributing to the charged fermion and Dirac neutrino masses. Subsequent analyses have shown that this Yukawa sector can explain small quark mixings and large leptonic mixings simultaneously  \cite{Bajc:2002iw,Fukuyama:2002ch,Goh:2003sy,Goh:2003hf,Bertolini:2004eq,Babu:2005ia,Bertolini:2006pe,Joshipura:2011nn,Altarelli:2013aqa,Dueck:2013gca,Bajc:2008dc,Fukuyama:2015kra,Fukuyama:2016vgi,Babu:2018tfi,Deppisch:2018flu}.  The model has 12 real parameters and 7 phases to fit 18 measured quantities including the neutrino oscillation parameters, leading to certain predictions.  In particular, the model prediction for the reactor neutrino angle was borne out by experiments \cite{DayaBay}.  

To complete the symmetry breaking, a {\bf 210} and a {\bf 126} Higgs fields need to be employed \cite{Aulakh:1982sw,Clark:1982ai,Aulakh:2003kg}. Such a model, with \{{\bf 210} + {\bf 126} + ${\bf \overline{126}}$ + {\bf 10}\} of Higgs fields, can separately be consistent with $SO(10)$ symmetry breaking and fermion mass generation.  However, when these two requirements are combined, the model does not fare well \cite{Aulakh:2005bd,Bajc:2005qe,Aulakh:2005mw,Bertolini:2006pe} for the following reason.  A fit to the neutrino oscillation parameters sets the overall right-handed neutrino mass scale, and the $(B-L)$ symmetry breaking scale $v_R$ to be $(10^{12}-10^{13})$ GeV.  The breaking of $SO(10)$ symmetry requires, on the other hand, $v_R\sim (10^{15}-10^{16})$ GeV.  If $v_R$ is chosen to be in the phenomenologically viable range of  $(10^{12}-10^{13})$ GeV, certain colored Higgs multiplets would acquire masses of order $v_R^2/M_{\rm GUT} \sim (10^8-10^9)$ GeV.  This would spoil perturbative unification of gauge couplings, making the model inconsistent.  This problem can be resolved, while maintaining the minimality of the Yukawa sector of Eq. (\ref{Yuk}), by the introduction of a {\bf 54} Higgs field, which cannot couple to the {\bf 16} fermions \cite{Babu:2018tfi}. In this case, $SO(10)$ symmetry can break down to $SU(3)_c \times SU(2)_L \times U(1)_Y \times U(1)_{B-L}$ at the GUT scale once the {\bf 54} and {\bf 210} Higgs fields acquire VEVs.  The $(B-L)$ symmetry is broken subsequently at a lower scale $v_R \simeq 10^{13}$ GeV when the singlet components of ${\bf 126} + {\bf \overline{126}}$ acquire VEVs.  No exotica survives below the GUT scale, except for the singlets from ${\bf 126} + {\bf \overline{126}}$ to facilitate $(B-L)$ symmetry breaking.  Consequently, gauge coupling unification is maintained as in the MSSM in this model.\footnote{An alternative approach is to add a {\bf 120} of Higgs field, which could be used to pair up some of the would-be light states.  In this case the Yukawa superpotential of Eq. (\ref{Yuk}) would have an additional coupling matrix \cite{Dutta:2004zh,Mohapatra:2018biy}.}

The model we present is a PQ symmetric extension of the aforementioned SUSY $SO(10)$ model of Ref. \cite{Babu:2018tfi} with the minimal Yukawa sector of Eq. (\ref{Yuk}).  The model thus contains ${\bf 210} + {\bf 126} + {\bf \overline{126}} + {\bf 10} + {\bf 54}$ of Higgs fields.  To implement the PQ symmetry, we also use an additional {\bf 10} and a set of $SO(10)$ singlet fields for breaking the PQ symmetry in the SUSY limit.  These fields and their PQ charges are listed in Table \ref{PQ-charge}.  The $U(1)_{\rm PQ}$ charges are chosen such that the Yukawa superpotential of Eq. (\ref{Yuk}) is allowed, and $SO(10)$ symmetry breaking can be achieved consistently.  Note that the fields that acquire GUT scale VEVs, viz., {\bf 210} and {\bf 54}, carry no PQ charge.

\FloatBarrier
\begin{table}[h!]
\centering
\footnotesize
\resizebox{1\textwidth}{!}{
\begin{tabular}{|c|c|c|c|c|c|c|c|c|c|c|}
\hline
Fields&$16_{F_i}$&$210_H$&$54_H$&$126_H$ &$\overline{126}_H$&$10_H$&$10^{\prime}_H$&$S_1$&$S_2$&$S_3$ \\ \hline
$U(1)_{\rm PQ}$ Charge&$-1$ & $0$ & $0$ & $-2$ & $+2$ & $+2$ & $-2$ & $-8$ &$+8$ & $+4$ \\ \hline
\end{tabular}
}
\caption{ PQ charge assignment of the fermions and the Higgs fields.}
\label{PQ-charge}
\end{table}

Without the PQ symmetry, SUSY $SO(10)$ models with the minimal Yukawa sector would require a mini-split SUSY spectrum \cite{Babu:2018tfi} to suppress the decay rate for $p \rightarrow \overline{\nu} K^+$ arising via dimension-5 operators mediated by the Higssinos. In the PQ symmetric model, these baryon number violating operators are induced only after the PQ symmetry is spontaneously broken, leading to a suppression factor of $(M_{\rm PQ}/M_{\rm GUT})^2$ in the decay rate.  Thus, neither the mini-split SUSY spectrum of Ref. \cite{Babu:2018tfi} with the minimal Yukawa sector, nor the  cancellation mechanism adopted in the color-triplet Higgs Yukawa couplings with an extended Yukawa sector that also includes couplings to a {\bf 120} of Higgs boson \cite{Dutta:2004zh,Mohapatra:2018biy}, is necessary.

To see the viability of the model and to understand the importance of introducing $10^{\prime}_H$-multiplet, we first write down  the superpotential involving the Higgs fields consistent with the  PQ charge assignment of Table \ref{PQ-charge}: 
\bea
W_{SO(10)}^{\rm PQ}&=&\frac{1}{2}m_1210_H^2+m_2\overline{126}_H126_H+\textcolor{red}{m^{\prime}_3}10_H 10^{\prime}_H+\frac{1}{2} m_5 54_H^2\non\\
&+&\lambda_1210_H^3+\lambda_2210_H\overline{126}_H126_H+\lambda_3126_H10_H210_H
+\textcolor{red}{\lambda_4^{\prime}}\overline{126}_H10_H^{\prime}210_H\non\\
&+&\lambda_854_H^3+\lambda_{10}54_H210_H^2+\textcolor{red}{\lambda_{13}^{\prime}}54_H10_H 10^{\prime}_H + \textcolor{red}{\lambda_5^{\prime}}10^{\prime\;2}_{H}S_3 .
\label{potential}
\eea

\noindent
In the next subsection we will discuss the breaking of PQ symmetry in the SUSY limit that involves $SO(10)$ singlet fields $S_{1,2,3}$. There we will see that all these singlet fields acquire VEVs which will be taken to be of order $(10^{11}-10^{12})$ GeV.  Among 
these fields, owing to our charge assignment, only $S_3$ has coupling with the $SO(10)$ non-singlet fields. With the superpotantial of Eq. (\ref{potential})
the mass spectrum of the SM non-singlet fields  can be found readily from the results given in Ref. \cite{Fukuyama:2004ps}.  Certain couplings are absent in our case however, due to the PQ symmetry. One must set
$m_3= \lambda_4= \lambda_{11}= \lambda_{12}= \lambda_{13}=0$ in the results of Ref. \cite{Fukuyama:2004ps}. Due to the presence of an additional $10^{\prime}_H$-multiplet, which contains $SU(2)_L$ doublet and $SU(3)_c$ triplet fields, the weak-doublet and color-triplet mass matrices get altered compared to the results of Ref. \cite{Fukuyama:2004ps}.  We present these matrices here, which play important roles in the fermion mass fit and dimension-5 baryon number violation.
The explicit form of the doublet  and triplet mass matrices would also help understand the need for the $10^{\prime}_H$-multiplet in the theory.   From the superpotential Eq. \eqref{potential}  we find the    
$SU(2)_L$ doublet mass matrix to be
\begin{align}\label{WD}
W^{mass}_D=\begin{pmatrix}
H_d&\overline{\Delta}_d&\Delta_d&\Phi_d&H^{\prime}_d
\end{pmatrix}
\mathcal{M}_D
\begin{pmatrix}
H_u& \Delta_u& \overline{\Delta}_u& \Phi_u& H^{\prime}_u
\end{pmatrix}^T ,
\end{align}

\noindent with
\begin{align}\label{doublet}
\mathcal{M}_D=\left(
\scalemath{0.85}{
\begin{array}{ccccc}
0&\frac{\lambda_3\Phi_2}{\sqrt{10}}-\frac{\lambda_3\Phi_3}{2\sqrt{5}}&0&0&m_3^{\prime}+\sqrt{3/5}\lambda^{\prime}_{13} E\\
0&m_2+\frac{\lambda_2\Phi_2}{15\sqrt{2}}-\frac{\lambda_2\Phi_3}{30}&0&0&\frac{\lambda_4^{\prime}\Phi_2}{\sqrt{10}}-\frac{\lambda_4^{\prime}\Phi_3}{2\sqrt{5}}\\
-\frac{\lambda_3\Phi_2}{\sqrt{10}}-\frac{\lambda_3\Phi_3}{2\sqrt{5}}&0&m_2+\frac{\lambda_2\Phi_2}{15\sqrt{2}}+\frac{\lambda_2\Phi_3}{30}&\frac{\lambda_2 \overline{v}_R}{10}&0\\
-\frac{\lambda_3 v_R}{\sqrt{5}}&0&\frac{\lambda_2 v_R}{10}&m_1+\frac{\lambda_1\Phi_2}{\sqrt{2}}+\frac{\lambda_1\Phi_3}{\sqrt{2}}-\frac{\sqrt{3}}{4\sqrt{5}} \lambda_{10} E&0\\
m_3^{\prime}+\sqrt{3/5}\lambda^{\prime}_{13} E&0&-\frac{\lambda_4^{\prime}\Phi_2}{\sqrt{10}}-\frac{\lambda_4^{\prime}\Phi_3}{2\sqrt{5}}&-\lambda^{\prime}_4 \frac{\overline{v}_R}{\sqrt{5}}&\lambda^{\prime}_5 \langle S_3\rangle
\end{array}
}
\right).
\end{align}

\noindent
Here the $\Phi_{1}= \langle (1,1,1) \rangle$, $\Phi_{2}= \langle (1,1,15) \rangle$ and $\Phi_{3}= \langle (1,3,15) \rangle$ are the VEVs
of $\Phi (210_H)$ and the $54_H$ VEV is $E= \langle (1,1,1) \rangle$ under the Pati-Salam group $SU(2)_{L}\times SU(2)_{R} \times SU(4)_{C}$ decomposition.  And  the color-triplet mass matrix is found to be
\begin{align}
W^{mass}_T=
\begin{pmatrix}
H_{\overline{c}}&\overline{\Delta}_{\overline{c}}&\Delta_{\overline{c}}
&\Delta^{\prime}_{\overline{c}}&\Phi_{\overline{c}}
&H^{\prime}_{\overline{c}}
\end{pmatrix}
\mathcal{M}_T
\begin{pmatrix}
H_c& \Delta_c& \overline{\Delta}_c & \overline{\Delta}_c^{\prime} & \Phi_c & H^{\prime}_c
\end{pmatrix}^T ,
\end{align}
 
\noindent with
\begin{align}\label{triplet}
\mathcal{M}_T=\left(
\scalemath{0.75}{
\begin{array}{cccccc}
0&\frac{\lambda_3\Phi_2}{\sqrt{30}}-\frac{\lambda_3\Phi_1}{\sqrt{10}}&0&0&0&m_3^{\prime}-\frac{2}{\sqrt{15}}\lambda^{\prime}_{13} E
\\
0&m_2&0&0&0&\frac{\lambda_4^{\prime}\Phi_2}{\sqrt{30}}-\frac{\lambda_4^{\prime}\Phi_1}{\sqrt{10}}
\\
-\frac{\lambda_3\Phi_1}{\sqrt{10}}-\frac{\lambda_3\Phi_2}{\sqrt{30}}&0&m_2&\frac{\lambda_2 \Phi_3}{15\sqrt{2}}&-\frac{\lambda_2 \overline{v_R}}{10 \sqrt{3}}&0
\\
-\sqrt{2/15}\lambda_3\Phi_3&0&\frac{\lambda_2\Phi_3}{15\sqrt{2}}&m_2+\frac{\lambda_2\Phi_1}{10\sqrt{6}}+\frac{\lambda_2\Phi_2}{30\sqrt{2}}&-\frac{\lambda_2\overline{v}R}{5\sqrt{6}}&0
\\
\frac{\lambda_3 v_R}{\sqrt{5}}&0&-\frac{\lambda_2 v_R}{10\sqrt{3}}&-\frac{\lambda_2 v_R}{5\sqrt{6}}&m_1+\frac{\lambda_1\Phi_1}{\sqrt{6}}+\frac{\lambda_1\Phi_2}{3\sqrt{2}}+\frac{2}{3}\lambda_1\Phi_3+\frac{1}{2\sqrt{15}} \lambda_{10} E&0
\\
m_3^{\prime}-\frac{2}{\sqrt{15}}\lambda^{\prime}_{13} E&0&-\frac{\lambda_4^{\prime}\Phi_1}{\sqrt{10}}-\frac{\lambda_4^{\prime}\Phi_2}{\sqrt{30}}&-\sqrt{\frac{2}{15}} \lambda^{\prime}_4 \Phi_3&\lambda^{\prime}_4 \frac{\overline{v}_R}{\sqrt{5}}&\lambda^{\prime}_5 \langle S_3\rangle
\end{array}
}
\right)  .
\end{align}

\noindent
It is evident from these matrices that if the $10^{\prime}_H$ field is not present, then the last row and the last column in both the mass matrices would be absent, which would result in one massless state in each sector. Electroweak symmetry breaking would generate mass to these states, which is however phenomenologically unacceptable since the light color-triplet would mediate rapid proton decay. One could avoid this by assigning $S_3$ a PQ charge such that the superpotential coupling $S_3 10^2_H$ is allowed.  In this case, the (1,1) entry will be nonzero in the doublet and triplet mass matrices.  The color triplet would then acquire a mass of order the PQ breaking scale, of order $10^{11}$ GeV, still leading to rapid proton decay.  To give large mass to the color triplet fields and avoid rapid proton decay, the simplest choice is to extend the Higgs sector by the addition of a $10^{\prime}_H$ with a PQ charge shown in Table \ref{PQ-charge}. Note that $10^{\prime}_H$ has no Yukawa coupling with the fermions owing to the PQ charge, so the Yukawa superpotential remains minimal as in Eq. (\ref{Yuk}).

Another important distinction between this PQ symmetric version and the conventional SUSY $SO(10)$ GUTs without PQ symmetry is that a second pair of Higgs doublets would have mass at the PQ scale. 
Note that in the PQ symmetric limit $\langle S_3\rangle \to 0$, the doublet mass matrix given in Eq. \eqref{doublet} would take a block-diagonal form:
\begin{align}\label{doublet-split} 
W^{mass}_D= &
\begin{pmatrix}
H^{\prime}_d&\Delta_d&\Phi_d
\end{pmatrix}
\mathcal{M}_{D_{I}}
\begin{pmatrix}
H_u& \overline{\Delta}_u& \Phi_u
\end{pmatrix}^T 
+
\begin{pmatrix}
H_d&\overline{\Delta}_d
\end{pmatrix}
\mathcal{M}_{D_{II}}
\begin{pmatrix}
H^{\prime}_u& \Delta_u
\end{pmatrix}^T 
+\lambda^{\prime}_5 H^{\prime}_dH^{\prime}_u \langle S_3\rangle.
\end{align} 

\noindent
Here $\mathcal{M}_{D_{I}}$ and $\mathcal{M}_{D_{II}}$ are sub-matrices of $\mathcal{M}_{D}$ of Eq. (\ref{doublet}), obtained by setting $\langle S_3\rangle = 0$.  It is clear that if one fine-tunes the $\mathcal{M}_{D_{I}}$ matrix   to keep a pair of Higgs doublet at the electroweak (EW) scale, the down-type quarks and charged leptons would remain massless since the MSSM $H_d$ field would be composed of a linear combination of ($H^{\prime}_d,\, \Delta_d,\, \Phi_d$) fields, with no component having Yukawa couplings with the down-type quarks and charge leptons.  Similarly, if the matrix $\mathcal{M}_{D_{II}}$ is fine-tuned to generate a weak scale doublet pair, 
$H_u$ of MSSM would be composed of ($H^{\prime}_u, \, \Delta_u$) fields, with neither component having Yukawa couplings with the up-type quarks.  It would be necessary to do two fine-tunings, one in each sector of $\mathcal{M}_{D_{I}}$ and $\mathcal{M}_{D_{II}}$, to generate both up-type and down-type quark masses, which would however result in two pairs of $(H_u,\,H_d)$ at the weak scale, ruining gauge coupling unification.   To resolve this issue,  mini fine-tunings are required to bring down two pairs of 
Higgs doublet to the PQ-scale -- one from $\mathcal{M}_{D_{I}}$ and one from $\mathcal{M}_{D_{II}}$ -- followed by  a large fine-tuning to make one combination of these two pairs at the weak scale. The weak scale doublets so obtained will have components from $(H_u,\,\overline{\Delta}_u)$ as well as $(H_d,\,\overline{\Delta}_d)$, leading to realistic masses for both up-type quarks, down-type quarks and charged leptons.  
Note that the mixing of these two pairs of Higgs doublets at the PQ scale is of order one, induced by the $\lambda_5' \langle S_3\rangle$ term, which is of the same order as the masses of the two pairs.  This mixing cannot be much smaller than unity, or else some of the fermion masses will be too small.

To identify the MSSM Higgs doublets, let us denote the light Higgs doublet pair emerging from $\mathcal{M}_{D_I}$ as $H^u_1+ H^d_1$  and those from  ($\mathcal{M}_{D_{II}}$) as $H^u_2+ H^d_2$. The remaining Higgs doublets will all have GUT scale masses, which we shall denote as $\tilde{H}$.  Then the states $H_u'$ and $H_d'$, contained in the $10'_H$, can be written as 
\begin{equation}
H_u' = \alpha H_2^u + \beta \tilde{H}_u,~~~ H_d' = \alpha_1 H_1^d + \beta_1 \tilde{H}_1^d + \gamma_1 \tilde{H}_2^d~
\end{equation}
where $(\alpha, \beta)$ and  $\alpha_i$ are elements of two unitary matrices obeying $|\alpha|^2+|\beta|^2 = |\alpha_1|^2+|\alpha_2|^2+|\alpha_3|^2 = 1$.  
Noting that $H_u'H_d' = \alpha \alpha_1 H_2^u H_1^d + ...$ with the $...$ containing GUT scale fields, the
superpotential relevant for the light MSSM Higgs doublets can be written as: 
\begin{align}\label{wwww}
W\supset \begin{pmatrix}
H^u_1&H^u_2
\end{pmatrix}
\begin{pmatrix}
\mu_{11}&0\\
\lambda^{\prime}\langle S_3\rangle \alpha \alpha_1&\mu_{22}
\end{pmatrix}
\begin{pmatrix}
H^d_1\\H^d_2
\end{pmatrix}.
\end{align}     
Here $\mu_{11}$ and $\mu_{22}$ are the effective masses of the doublet pairs from $\mathcal{M}_{D_I}$ and
$\mathcal{M}_{D_{II}}$ respectively, taken to be at the PQ scale.  
Furthermore, the Yukawa superpotential with the doublets at the PQ scale is given by:
\begin{align}\label{YY}
W_Y\supset y_uQu^cH^u_1+y_dQd^cH^d_2+y_eLe^cH^d_2+y_{\nu}L\nu^cH^u_1.
\end{align}  

\noindent
Here the matrix elements $\mu_{ij}$ in Eq. \eqref{wwww} and the effective Yukawa couplings $y_i$ in Eq. \eqref{YY} can be computed straightforwardly. From the Yukawa couplings in Eq. \eqref{YY} it is clear that $H^u_1$ and $H^d_2$ must be part of light MSSM states to generate realistic fermion mass. This can be achieved by performing a large fine-tuning by setting the determinant of the mass matrix in Eq. (\ref{wwww}) equal to zero, which reuires $\mu_{11}\mu_{22}=0$. If $\mu_{22}=0$ is chosen, the MSSM Higgs doublet pair will be identified as: 
\begin{align}
H^{MSSM}_d= H^d_2,\;\;\;  H^{MSSM}_u= \frac{\lambda^{\prime}\langle S_3\rangle \alpha \alpha_1 H^u_1 - \mu_{11} H^u_2}{\left[  \mu^2_{11} +(\lambda^{\prime}\langle S_3\rangle \alpha \alpha_1)^2  \right]^{1/2}}.  
\end{align}

\noindent Furthermore, if $\mu_{11}< \left( \lambda^{\prime}\langle S_3\rangle \alpha \alpha_1\right)$ is assumed, then $H^{MSSM}_u\simeq H^u_1$, as expected. It can be readily verified that this fine-tuning does not lead to any light color triplet.  These arguments show the consistency of the fine-tuning framework, as well as the necessity of having a second Higgs doublet pair at the PQ scale.

For numerical computations performed later in the text, we use exact relations by diagonalizing the $5\times 5$ doublet matrix.  This matrix $\mathcal{M}_D$ of Eq. \eqref{doublet} contains a pair of doublets at the PQ scale and another pair at the weak scale, which is identified as the MSSM Higgs fields. The needed tunings should be achieved while making sure that all the eigenvalues of the color  triplet mass matrix $\mathcal{M}_T$ lie at the GUT scale to avoid rapid proton decay. The MSSM Higgs  $H^{MSSM}_u$ and $H^{MSSM}_d$ are mixtures of the doublets coming from $10_H$, $10^{\prime}_H$, $126_H$, $\overline{126}_H$ and $210_H$  multiplets denoted as:
\begin{align}
&H^{MSSM}_u=N_u (H_u+ p_2 \Delta_u+ p_3 \overline{\Delta}_u+p_4\Phi_u+p_5H^{\prime}_u),\\
&H^{MSSM}_d=N_d (H_d+ q_2 \overline{\Delta}_d+ q_3 \Delta_d+q_4\Phi_d+q_5H^{\prime}_d).
\end{align} 

\noindent
The normalization factors are defined as:
$N_u= \left( 1+\sum_{i=2}^5|p_i|^2 \right)^{-1/2}$ and $N_d= \left( 1+\sum_{i=2}^5|q_i|^2 \right)^{-1/2}$. 
The  expressions for $p_i$ and $q_i$ can be found in a straightforward way from the left and the right eigenvectors corresponding to the zero eigenvalue of Eq. \eqref{doublet}. We compute these numerically in our estimate of proton decay rate.  Two important parameters, denoted as $r$ and $s$, which appear in the fermion mass fits are
determined in terms of these superpotential parameters:
\begin{align}\label{rs}
r=N_u/N_d,\;\;\; s=p^{\ast}_3/q^{\ast}_2 .
\end{align}
The best fit values of these parameters from fermion mass spectrum are given in Eq. (\ref{rs-values}).

\subsection{Symmetry breaking chain} 
In our set-up the symmetry breaking proceeds in three steps (Chain A):
\begin{eqnarray}\label{chain}
SO(10)\times U(1)_{\rm PQ} &\xrightarrow{\langle 54_H \rangle,\,\langle 210_H \rangle}& SU(3)_c \times SU(2)_L \times U(1)_Y  \times U(1)_{B-L} \times U(1)_{\rm PQ} \nonumber \\
&\xrightarrow{\langle 126_H \rangle,\,\langle \overline{126}_H \rangle,\, \langle S_i \rangle}& SU(3)_c \times SU(2)_L \times U(1)_Y \nonumber \\
&\xrightarrow{\langle 10_H \rangle,\, \langle 10^{\prime}_H \rangle}, ... & SU(3)_c \times U(1)_{\rm em}
\end{eqnarray}

\noindent
The first step of symmetry breaking occurs at $M_{\rm GUT}\approx 2 \times 10^{16}$ GeV, which leaves the rank of $SO(10)$ group intact.  At the next step, the VEVs of $126_H+\overline{126}_H$ multiplets   break the   $U(1)_{B-L}$ symmetry spontaneously.  And in the final step, Higgs doublets from the $10, 10'$ etc break the electroweak symmetry.  We assume this symmetry breaking chain in order to achieve a realistic fit for the fermion masses and mixings, while preserving the successful gauge coupling unification realized in the MSSM. Note that the  $U(1)_{B-L}$ breaking scale is determined to be $v_R\sim 10^{12}$ GeV $\ll M_{\rm GUT}$  from a fit to neutrino oscillation data with the minimal Yukawa couplings of Eq. \eqref{Yuk}.  Hence, decoupling of the $B-L$ scale from the GUT scale is demanded by phenomenology.  We shall discuss below other possible breaking chains and show that these scenarios are either not viable, or would require large threshold corrections to maintain gauge coupling unification.  In this discussion we focus only on the gauge symmetry; the breaking of the global $U(1)_{\rm PQ}$ symmetry will be addressed in the next subsection.

\begin{itemize}
\item \underline{Chain B:} 
Let us consider the case with $\langle 54_H\rangle = M_{\rm GUT}\gg \langle 210_H\rangle$, in which case the symmetry breaking scheme would be $SO(10)\xrightarrow{M_{\rm GUT}}  SU(2)_L\times SU(2)_R \times SU(4)_C\times D \equiv G_{224} \times D$.  This is the Pati-Salam gauge symmetry along with a  discrete $D$-parity \cite{Kibble:1982dd, Chang:1983fu}, which interchanges $SU(2)_L$ and $SU(2)_R$. Then a second intermediate scale  $M_I \ll M_{\rm GUT}$ would be needed (with the possibility that  $M_I \sim v_R$) to break the intermediate  $G_{224} \times D$ symmetry down to our desired  $SM\times U(1)_{B-L}$ symmetry. This symmetry breaking can be realized when the $(1,1,15)$ and $(1,3,15)$ multiplets from $210_H$ (under $G_{224}$) acquire  VEVs at the $M_I$ scale. Furthermore, the generation of realistic fermion masses and mixings also requires $(2,2,1)\subset 10_H$ and $(2,2,15)\subset \overline{126}_H$ to reside at the intermediate scale.  Note that, due to the $D$-parity, the parity partner of $(1,3,15)$, viz., $(3,1,15)$ multiplet, would also remain light at the scale $M_I$ in this scenario. If $M_I \approx v_R$, then due to the presence of too many particles at the scale of $\sim 10^{12}$ GeV, perturbative unification of the gauge couplings would be spoiled. A scenario with of $v_R \ll M_X < M_{\rm GUT}$ could be viable; however, in this case large threshold corrections would be needed to maintain gauge coupling unification.

\item \underline{Chain C:}
One can also have the possibility where $\langle 210_H \rangle = M_{\rm GUT}> \langle 54_H\rangle$.
Within this scenario, few different symmetry breaking chains may be realized (chains C, D, E). If the singlet components of the $210_H$ Higgs all acquire VEVs following the relation $\sqrt{6}\langle (1,1,1)\rangle = \sqrt{2}\langle (1,1,15)\rangle =\pm \langle (1,3,15)\rangle$, then $SO(10)\xrightarrow{M_{\rm GUT}} SU(5)\times U(1)$ symmetric vacuum appears.  
This symmetry needs to be further broken by the VEVs of $54_H$ to give rise to SM $\times U(1)_{B-L}$ symmetry at a scale $M_I \ll M_{\rm GUT}$.
This symmetry breaking chain is certainly not viable, since it would lead to rapid proton decay induced by the $SU(5)$ gauge bosons, resulting in a lifetime that is in conflict with experimental limits.

\item \underline{Chain D:}
If only the $(1,1,15)\subset 210_H$ field acquires a VEV at the GUT scale, then symmetry breaking proceeds via the left-right symmetric vacuum: $SO(10)\rightarrow SU(3)_c\times  SU(2)_L\times SU(2)_R\times U(1)_{B-L}\times D$. This symmetry may be broken at an intermediate scale $M_I$ when $(1,1,3,0)$ multiplet (under the LR symmetric gauge symmetry) from $210_H$ acquires VEV. In this scenario, bi-doublets $(1,2,2,0)$ from $10_H$ and $\overline{126}_H$  should be at the intermediate scale to generate correct fermion mass.  Due to $D$-parity, $(1,3,1,0)$ also would survive to this intermediate scale.  As a consequence of these new fields at $M_I$, gauge coupling unification would be spoiled, unless large threshold corrections from GUT scale particles are invoked.  If on the other hand  $v_R \ll M_I < M_{\rm GUT}$ is assumed, somewhat smaller  threshold correction would be sufficient to maintain coupling unification.

\item \underline{Chain E:}
In this last scenario, all three multiplets  $(1,1,1)$, $(1,1,15)$, $(1,3,15)$ $\subset 210_H$  acquire nonzero VEVs at the GUT scale  leading to the symmetry breaking chain $SO(10)\rightarrow SU(3)_c \times SU(2)_L\times U(1)_R\times U(1)_{B-L}$.  The $U(1)_R \times U(1)_{B-L}$ symmetry is left unbroken by the $54_H$ field, but is broken by the $\overline{126}_H + 126_H$ fields.  This scenario would lead to pseudo-Goldstone Higgs multiplets carrying SM quantum numbers with masses of order $\sim v^2_R/M_{\rm GUT}\sim 10^{10}$ GeV.  Perurbative unification of gauge couplings would be spoiled in this scenario.

\end{itemize}

From the above analysis it is clear that  our choice of the symmetry breaking chain given in  Eq. \eqref{chain} (Chain A) is the only one that generates a realistic fermion mass spectrum, while being consistent with gauge coupling unification without relying on excessive threshold effects.

\subsection{PQ symmetry breaking via singlet fields}\label{PQ-breaking}
Now we focus on the  PQ symmetry which  also needs to be broken around this intermediate scale, $M_{\rm PQ}\approx v_R$, although these two symmetries are a priori unrelated. Note that the SM singlet fields $\sigma+\overline{\sigma}$ from $  126_H+\overline{126}_H$ carrying two units of $B-L$ charge that break the gauged $U(1)_{B-L}$ cannot simultaneously break the global  $U(1)_{\rm PQ}$ symmetry. 
Additional SM singlet fields carrying non-zero PQ charge are necessary for this purpose.  The two sets of fields jointly  break $U(1)_{B-L}\times U(1)_{\rm PQ}$ symmetry completely. If only the SM singlets from $\overline{126}_H + 126_H$ were to acquire VEVs, one linear combination of $B-L$ and $PQ$ symmetries  would remain unbroken. To see the unbroken global symmetry we first list the the $SO(10)\rightarrow SU(5)\times U(1)_X$ decomposition of all the fields:
\begin{align}
&10=5_{2}+\overline{5}_{-2}
\\
&16=1_{-5}+\overline{5}_3+10_{-1}
\\
&54=15_{4}+\overline{15}_{-4}+24_0
\\
&126=1_{-10}+\overline{5}_{-2}+10_{-6}+\overline{15}_{6}+45_{2}+\overline{50}_{-2}
\\
&210=1_{0}+5_{-8}+\overline{5}_{8}+10_{4}+\overline{10}_{-4}+24_{0}+40_{-4}+\overline{40}_{4}+75_{0}.
\end{align}

\noindent From these, we find that the $54_H, 126_H, \overline{126}_H$ and $210_H$ VEVs leave a global symmetry, $U(1)_{PQ^{\prime}}$ unbroken, given by $PQ^{\prime}= (5 PQ-X)/4$. It is because all the fields that get VEVs, including $\sigma+\overline{\sigma}$ carry zero charge under $PQ^{\prime}$. The charges of the isospin doublets and the color triplets  under this $PQ'$ symmetry are  shown in Table  \ref{BLPQ}.  Since the original PQ symmetry and the $U(1)_X$ symmetry commute with $SU(5)$, the $PQ'$ charges are the same for all members of a given $SU(5)$ multiplet.  The existence of the unbroken global $PQ'$ symmetry shows that $SO(10)$ singlet fields are necessary for consistent symmetry breaking. Without these singlets, the $PQ'$ symmetry would be broken by Higgs doublets, leading to weak scale axion excluded by experiments.

\FloatBarrier
\begin{table}[th!]

\centering
\footnotesize 
\resizebox{0.9\textwidth}{!}{
\begin{tabular}{|c|c|c|c|c|c|c|c|c|c|c|c|c|c|}
\hline

fields&$H_c$&$H_{\overline{c}}$&$\overline{\Delta}_c$&$\overline{\Delta}_{\overline{c}}$&$\Delta_c$&$\Delta_{\overline{c}}$&$\Phi_c$&$\Phi_{\overline{c}}$&$H^{\prime}_c$&$H^{\prime}_{\overline{c}}$ &$\overline{\Delta}^{\prime}_c$ & $\Delta^{\prime}_{\overline{c}}$ \\ \hline

\pbox{10cm}{\vspace{2pt}$U(1)_{PQ^{\prime}}$\\ charge\vspace{2pt}} & $2$ &$3$&$2$&$3$&$-3$&$-2$&$2$&$-2$&$-3$&$-2$&$2$&$-2$\\ \hline

\end{tabular} 
} 
\caption{ $U(1)_{PQ^{\prime}}$ charges of the color-triplets with  $PQ^{\prime}=(5 PQ-X)/4$. Note that except $\overline{\Delta}^{\prime}_c$ and $\Delta^{\prime}_{\overline{c}}$ multiplets, there is an associated   isospin-doublet partner originating from the same Higgs field and carrying identical  $U(1)_{PQ^{\prime}}$ charge as that of the color-triplet.     }\label{BLPQ}
\end{table}

To break the PQ symmetry consistently,  we introduce three singlet fields $S_{1,2,3}$ that carry non-trivial charges under $U(1)_{\rm PQ}$. Their PQ charges are listed in Table  \ref{PQ-charge}. From various experimental bounds, the PQ breaking scale $f_a$ is restricted to be within the range  $10^{10}$ GeV $\lesssim f_a \lesssim 10^{12}$ GeV. The superpotantial consisting of the $SO(10)$ singlet fields is 
given by:\footnote{An alternative choice is $W_S=\kappa S_3(S_1 S_2 - M_{\rm PQ}^2)$ with $S_3$ having zero charge and $S_1$ and $S_2$ carrying equal and opposite charges.} 
\begin{align}\label{singlet}
W_S=M_SS_1S_2+\kappa S_1 S_3^2 .
\end{align}

\noindent In the SUSY preserving limit all the $F$-terms must vanish. They are given by
\begin{equation}
F_{S_1}=M_S S_2+\kappa S^2_3,~~~F_{S_2}=M_SS_1,~~~F_{S_3}=2\kappa S_1 S_3~.
\end{equation}
This sets $\langle S_1 \rangle = 0$, and one combination of $\langle S_2 \rangle$ and $\langle S_3 \rangle$ is fixed. The undetermined VEV leads to a flat direction in $S_{2,3}$. This flat direction is lifted once soft SUSY breaking terms are included. The full potential, including soft SUSY breaking terms of the singlet sector is given by:
\begin{align}
&V=V_{soft}+|F_{S_1}|^2+|F_{S_2}|^2+|F_{S_3}|^2 ,\label{PQ1}\\
&V_{soft}=\{ B_S M_S S_1 S_2 + A_{\kappa} \kappa S_1 S^2_3 + h.c.  \} + m^2_1 |S_1|^2+ m^2_2 |S_2|^2+ m^2_3 |S_3|^2 \label{PQ2}. 
\end{align}

\noindent
Straightforward calculation shows that including soft SUSY breaking, all VEVs of the singlet fields are fixed. We find (treating SUSY breaking terms perturbatively)
\begin{align}
&\langle S_1\rangle =\frac{\kappa^{\ast}{\langle S_3^{\ast}\rangle}^{2}(B^{\ast}_S-A^{\ast}_{\kappa})}{|M_S|^2+4|\kappa|^2|\langle S_3\rangle|^2},\;\; 
\langle S_2\rangle=-\frac{\kappa}{M_S}\langle S_3\rangle^2+\delta  ,
\\
&A|\langle S_3\rangle^2|^3+B|\langle S_3\rangle^2|^2+C|\langle S_3\rangle^2|+D=0 .
\end{align}

\noindent
Here we have defined
\begin{align}
&A=32|\kappa|^6m^2_2,\;\; D=-m^2_3|M_S|^6,\;\; \delta=\frac{\kappa m^2_2 \langle S_3 \rangle^2}{|M_S|^2}-\frac{B_S^{\ast}\kappa \langle S_3\rangle^2(B_S-A_{\kappa})}{|M_S|^2+4|\kappa|^2|\langle S_3\rangle|^2} ,
\\
&B=16|\kappa|^4m^2_2|M_S|^2-4|\kappa|^4|M_S|^2|A_{\kappa}-B_S|^2-16|\kappa|^4m^2_3|M_S|^2 ,
\\
&C=2|\kappa|^2|M_S|^4m^2_2-2|\kappa|^2|M_S|^4|A_{\kappa}-B_S|^2-8|\kappa|^2|M_S|^4m^2_3 .
\end{align}

\noindent  This shows that the singlet sector can consistently break $U(1)_{\rm PQ}$ symmetry. All three singlets acquire masses of order the PQ scale. We shall use the VEV of $S_3$ as an independent parameter for our fermion mass fit and proton decay calculations, assuming that $\langle S_3 \rangle \sim  10^{11}$ GeV. 
The VEVs of $126+\overline{126}$ multiplets   along with the  singlets carrying non-zero PQ charge together  break the symmetry in such a way that a discrete $\mathcal{Z}_2$ symmetry remains unbroken. This is the well known $\mathcal{Z}_2$ subgroup that is identified as the R-parity which stabilize the dark matter in this setup. This R-parity can be identified with the well known matter parity $P_M=(-1)^{3(B-L)}$ under which all the fermions are odd and all the scalars are even.

\subsection{PQ symmetry and the domain wall problem}
Here we briefly discuss the cosmological consequences of the PQ symmetry breaking.  It is well known that  spontaneous breaking of the PQ symmetry leads to $N_{DW}$ distinct degenerate vacua in axion models due to a residual discrete $\mathcal{Z}_{N}$ symmetry. This $\mathcal{Z}_{N}$ symmetry,  which is a subgroup of $U(1)_{\rm PQ}$, is left unbroken by non-perturbative QCD effects. In QCD, the vacuum has non-trivial structure and for $N_{f}$ number of flavors has a global $SU(N_f)\times SU(N_f)\times U(1)_{V}\times U(1)_A$ symmetry. Of this global symmetry, the axial $U(1)_A$ is anomalous and is broken by the QCD instanton effects \cite{tHooft:1976rip}  down to a discrete symmetry $\mathcal{Z}_N$, with $N=2 N_g$ ($N_g$ is the number of quark generations). Topological objects called domain walls \cite{Kibble:1976sj} are formed as a consequence of this non-trivial structure of the QCD vacuum \cite{Sikivie:1982qv}.  
The associated  domain wall number, $N_{DW}$,  can be  computed  from the $[SU(3)_C]^2\times U(1)_{PQ}$ anomaly coefficient \cite{Kim:1986ax}.  

For clarity of discussion, let us assume that the $B-L$ symmetry breaking scale $v_R$ is slightly above the PQ symmetry breaking scale.  Once the singlet components of $\overline{126}_H$ and $126_H$ acquire VEVs, the $SO(10) \times U(1)_{PQ'}$ symmetry breaks down to MSSM $\times U(1)_{PQ'}$.  
As discussed in Sec. \ref{PQ-breaking}, the matter fields at this scale  
consist of the SM fermions, two-pairs of Higgs doublets, and three MSSM singlet superfields. 
The charges of the fermion fields under $U(1)_{PQ'}$ are listed in Table \ref{PQ-fermion}, and those of the singlet and doublet scalars are listed in Table \ref{PQ-singlet}.  
The superpotential involving these fields is given by
\begin{align}\label{WW}
W\supset \mu_{11}H^u_1H^d_1+\mu_{22}H^u_2H^d_2+\lambda_{21} S_3H^u_2H^d_1 + M_S S_1 S_2 + \kappa S_1 S^2_3.
\end{align}     
Note that the Lagrangian, which includes SUSY breaking terms, would contain bilinear and trilinear scalar couplings analogous to the superpotential terms given above.

\FloatBarrier
\begin{table}[th!]
\centering
\footnotesize 
\resizebox{0.5\textwidth}{!}{
\begin{tabular}{|c|c|c|c|c|c|c|c|c|c|c|c|}
\hline

fields&$Q$&$L$&$u^c$&$d^c$&$e^c$&$\nu^c$  \\ \hline

\pbox{10cm}{\vspace{2pt}$U(1)_{PQ^{\prime}}$\\ charge\vspace{2pt}} & 
$-1$&$-2$&$-1$&$-2$&$-1$&$0$ \\ \hline

\end{tabular} 
} 
\caption{ $U(1)_{PQ^{\prime}}$ charges of the fermions with  $PQ^{\prime}=(5 PQ-X)/4$.      }\label{PQ-fermion}
\end{table}

\FloatBarrier
\begin{table}[th!]
\centering
\footnotesize 
\resizebox{0.6\textwidth}{!}{
\begin{tabular}{|c|c|c|c|c|c|c|c|c|c|c|c|}
\hline

fields&$H^u_1$&$H^u_2$&$H^d_1$&$H^d_2$&$S_1$&$S_2$&$S_3$  \\ \hline

\pbox{10cm}{\vspace{2pt}$U(1)_{PQ^{\prime}}$\\ charge\vspace{2pt}} & 
$2$&$-3$&$-2$&$3$&$-10$&$10$&$5$ \\ \hline

\end{tabular} 
} 
\caption{ $U(1)_{PQ^{\prime}}$ charges of the SM singlets with  $PQ^{\prime}=(5 PQ-X)/4$.      }\label{PQ-singlet}
\end{table}

We now proceed to compute the domain wall number following Ref. \cite{ringwald}. The symmetry left unbroken by the QCD instanton effect is $\mathcal{Z}_N$. However, $N$ is not necessarily the number of domain walls, since  the transformation $\phi_i\to \phi_i +2\pi v_i$ is trivial when the fields are expressed  in the exponential parameterization. To determine the domain wall number, we write the fields appearing in Eq. \eqref{WW} as:
\begin{align}
&H^{u,d}_{1,2}=\frac{\rho^{u,d}_{1,2}+v^{u,d}_{1,2}}{\sqrt{2}} e^{i\phi^{u,d}_{1,2}/v^{u,d}_{1,2}},  \;\;\;
S_i=\frac{\rho_{i}+v_{i}}{\sqrt{2}} e^{i\phi_{i}/v_{i}}. 
\end{align}

We first identify the axion field which should be orthogonal to the massive fields, as well as the Goldstone field eaten up by the $Z$ boson.  Eq. \eqref{WW} and its soft SUSY breaking counterparts contain the following relevant terms:
\begin{align}\label{WWW}
{\cal L}\supset &
\frac{1}{2}v^u_1v^d_1e^{i(\phi^u_1/v^u_1+\phi^d_1/v^d_1)}
+\frac{1}{2}v^u_2v^d_2e^{i(\phi^u_2/v^u_2+\phi^d_2/v^d_2)}
+\frac{1}{2\sqrt{2}}v^u_2v^d_1v_3e^{i(\phi^u_2/v^u_2+\phi^d_1/v^d_1+\phi_3/v_3)}
\nonumber \\
&+\frac{1}{2}v_1v_2e^{i(\phi_1/v_1+\phi_2/v_2)}
+\frac{1}{2\sqrt{2}}v_1v_3^2e^{i(\phi_1/v_1+2\phi_3/v_3)}.
\end{align}
The Goldstone boson eaten up by the $Z$ boson is the combination 
\begin{align}
G\propto (v^u_1\phi^u_1-v^d_1\phi^d_1+v^u_2\phi^u_2-v^d_2\phi^d_2).
\end{align}
The axion field, which should be orthogonal to these fields is then found to be

\begin{align}
A=\frac{1}{f_{PQ}}\left( 2v_1\phi_1-2v_2\phi_2-v_3\phi_3+x(v^u_2\phi^u_2-v^d_2\phi^d_2)+(1-x)(v^d_1\phi^d_1-v^u_1\phi^u_1)  \right)     
\end{align}
\noindent where we have defined  $x$ as
\begin{align}
&x\equiv \frac{{v^u_1}^2+{v^d_1}^2}{v^2},\;\textrm{with}\;\; \;\;\; v^2\equiv {v^u_1}^2+{v^d_1}^2+{v^u_2}^2+{v^d_2}^2.
\end{align}
Note that $0 \leq x \leq 1$.  The axion decay constant is identified as
\begin{equation}
f_{PQ}^2=V^2 + x(1-x)v^2,  \;\;\;{\rm where}~~~
V^2\equiv 4(v_1^2+v_2^2)+v_3^2.  
\end{equation}

It is convenient to define the axion field as:
\begin{align}
&A=\frac{1}{f_{PQ}}\sum_i q_iv_i\phi_i,
\end{align}
\noindent in which case the charges $q_i$ are found to be: $q_1=2, q_2=-2$, $q_3=-1, q^u_1=x-1$, $q^d_1=1-x$, $q^u_2=x$, $q^d_2=-x$ and from Eq. \eqref{YY} they obey the relations:
\begin{align}
&q_Q+q_{u^c}=1-x,\;\;\;q_Q+q_{d^c}=x,\;\;\;
q_L+q_{e^c}=x,\;\;\;q_L+q_{\nu^c}=1-x.
\end{align}
\noindent This gives for the integer $N$ in the unbroken $Z_{N}$ for our model to be
\begin{align}
N\equiv N_g\left( 2q_Q+q_{u^c}+q_{d^c} \right)=N_g.    
\end{align}

\noindent Then the domain wall number is given by \cite{ringwald}:
\begin{align}
N_{DW}=\textrm{minimum integer}\left( N  \sum_i \frac{n_iq_iv^2_i}{f^2_{PQ}}  \right),\;\;\;n_i  \in \mathbb{Z}.
\end{align}

\noindent
After a straightforward algebra we find:
\begin{align}
&\sum_i \frac{n_iq_iv^2_i}{f^2_{PQ}} =\frac{(n_1-n_2-4n_3)\frac{V^2}{8} +x(1-x)v^2(n^d_1-n^d_2)}{V^2 + x(1-x)v^2} \nonumber\\
&+\frac{\frac{1}{2}(n_1-n_2+4n_3)(v^2_1+v^2_2-\frac{v^2_3}{4})+ (n_1+n_2)(v^2_1-v^2_2)-{v^u_1}^2(1-x)(n^u_1+n^d_1)+ {v^u_2}^2x(n^u_2+n^d_2)}{V^2 + x(1-x)v^2}.
\end{align}
\noindent To get integer value for $N_{DW}$, the $n_i$ are strongly restricted, as the monomials in the numerator and denominator should be proportional.  We find
\begin{align}
n_1=-2n_3, n_2=2n_3,n^d_1=-n^u_1, n^d_2=-n^u_2.    
\end{align}
\noindent Putting these relations back in the original expression gives:
\begin{align}
\sum_i \frac{n_iq_iv^2_i}{f^2_{PQ}} =\frac{-n_3V^2+x(1-x)v^2(n^u_2-n^u_1)}{V^2 + x(1-x)v^2}.
\end{align}
\noindent This restricts $n_3=n^u_1-n^u_2$, and  substituting this back gives:
\begin{align}
N_{DW}=\textrm{minimum integer}\left( N  \sum_i \frac{n_iq_iv^2_i}{f^2_{PQ}}  \right)= min\left( N (n^u_2-n^u_1) \right)= N=N_g.
\end{align}
\noindent So the number of domain walls we have in our theory is $N_g=3$.

The axion acquires a mass  due to non-perturbative QCD effect given by \cite{Abbott:1982af, Preskill:1982cy, Dine:1982ah}:
\begin{align}
m_a\approx N \left( 6\times 10^{-6} eV \right)  \left( \frac{10^{12} GeV}{f_{PQ}} \right).
\end{align} 

\noindent 
Note that, as a result of the  PQ phase transition around the temperature of order $T_{PQ}\approx f_{PQ}$  one-dimensional topological
defects called axionic strings are formed \cite{Kibble:1976sj}. These axion strings radiate axions in between temperatures $T_{PQ}$ and $T_{QCD}$ and at $T_{QCD}$ each string becomes the boundry of $N_{DW}$ domain walls  and form stable topological defects that are known as the string-domain wall networks \cite{Davis:1985pt, Davis:1986xc, Harari:1987ht}.   It should be pointed out that for $N_{DW}=1$, the string-wall network is unstable, decaying into axions soon after their fomration \cite{Vilenkin:1982ks, Vilenkin:1986ku}. In this case there is no cosmological problem. On the other hand, for $N_{DW}\geq 2$ these networks are quite stable. Such stable topological objects are disastrous  and would create a serious problem because energy density in the axion domain walls would dominate the universe soon after their formation and overclose the universe.

We assume that the axionic string-wall network problem is solved by having inflation with reheating temperature less than the PQ phase transition temperature \cite{Pi:1984pv}. When the PQ symmetry is broken before or during the inflation, each domain with a specific value of PQ phase $\theta_a$ expands to a size larger than the size of the present observable universe  resulting in no topological defects.   The value of the axion field becomes homogeneous throughout the whole observable universe as a result of the exponential cosmic expansion during the inflationary time.  However, further complexity arises since 
 the axion field obtains  fluctuations during inflation.
When the axions acquire mass at the QCD scale these fluctuations lead to  isocurvature density perturbations  \cite{Axenides:1983hj, Seckel:1985tj, Linde:1985yf, Linde:1990yj, Turner:1990uz, Linde:1991km, Lyth:1991ub}.
Observation of the cosmic microwave background (CMB) puts stringent constraints on the amplitude of these isocurvature perturbations.
Solutions of these cosmological problems depend on the details of the inflationary model. 
 By performing lattice simulations, it has been shown  that the domain wall problem as well as the isocurvature perturbation problem  can be solved \cite{Kawasaki:2013iha} simultaneously in general inflationary models that allows intermediate  PQ symmetry breaking scale. These numerical computations also show that chaotic inflation can solve both these problems together, although this requires the  initial  misalignment  angle $\theta_a$  to be  somewhat small  $\lesssim O(10^{-2})$ and axion cannot be the main component of the dark matter in the universe.  Similar lattice simulations have been carried out recently   within the supersymmetric framework assuming chaotic inflationary scenario in \cite{Kawasaki:2017kkr} where it is shown that an intermediate scale PQ symmetry  breaking, such as the one needed in our model, is completely consistent and devoid of all cosmological problems discussed above.

\subsection{Suppression of proton decay rate}
\begin{figure}[h!]
\begin{center}
\includegraphics[width=12.cm]{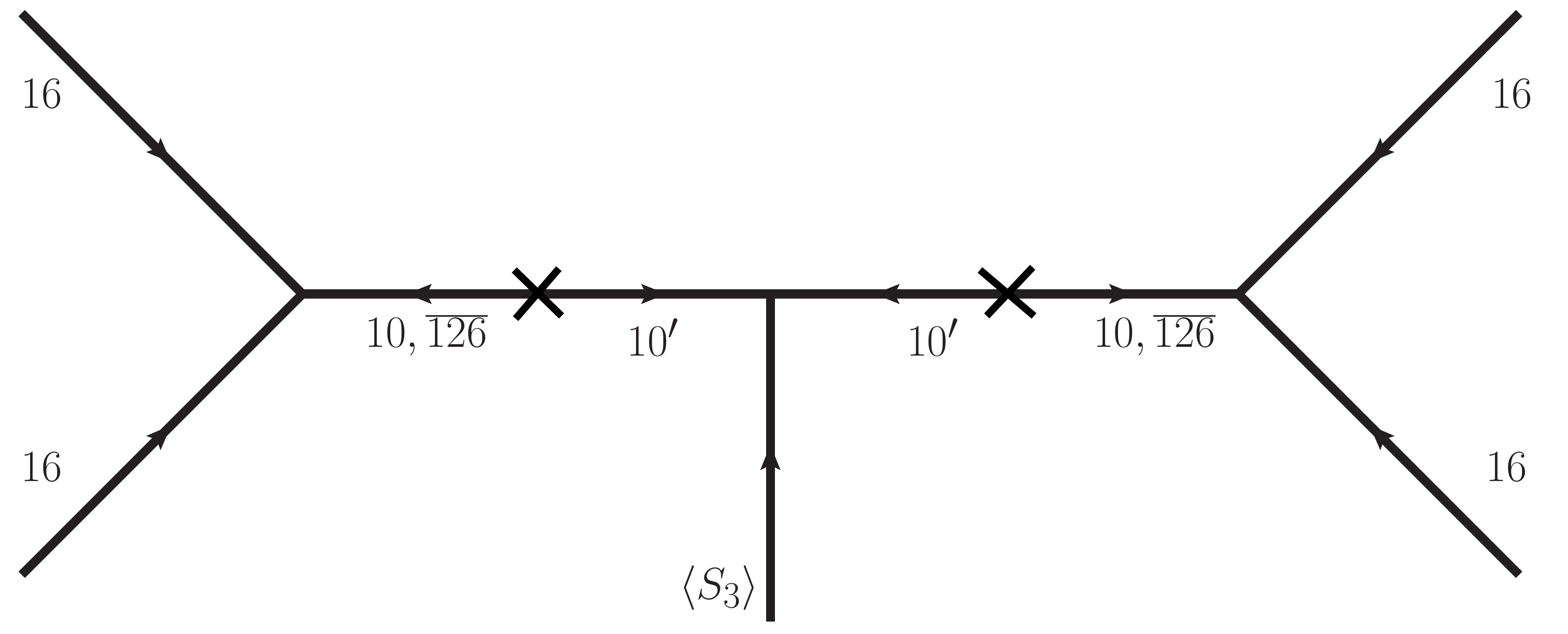}
\end{center}
\caption{ Feynman diagram responsible for proton decay in our theory. This diagram contains four different diagrams involving Yukawa couplings of color triplet fields from either the $10_H$ or $\overline{126}_H$ Higgs multiplets.}
\label{protondecay}
\end{figure}

The Peccei-Quinn symmetry present in the model leads to a strong suppression of proton decay rate induced by the color triplet Higgsinos.   
We now turn to this issue. In SUSY GUTs the $d=5$ baryon number violating operators induced by color triplet Higgsinos are more dominant compared to the $d=6$ operators mediated by the gauge bosons \cite{Weinberg:1981wj,Sakai:1981pk}. 
 As a result, the most dominant modes of proton decay are typically $p\to \overline{\nu} K^+$ and $p\to \overline{\nu} \pi^+$.  
The   Feynman diagram responsible for $d=5$ proton decay operators  mediated by the color-triplets in our PQ symmetric SUSY $SO(10)$ is shown in Fig \ref{protondecay}. Note that the $d=5$ superpotential operator $16_i 16_j 16_k 16_l$ is not invariant under PQ symmetry.  The net charge $-4$ of this operator is compensated by charge $+4$ of $S_3$ field that breaks PQ symmetry. The singlet sector superpotential allows for $\langle S_3\rangle \sim 10^{11}$ GeV, consistent with constraints from axion searches.  The mixing term $54_H10_H10^{\prime}_H$ present in the superpotential  Eq. \eqref{potential} allows for all color triplets to acquire GUT scale masses, $M_T\sim M_{\rm GUT}$. Then the $d=5$ effective  baryon number violating superpotential will take the form
\begin{align}
W_5\sim \left( \lambda^{\prime}_5\langle S_3 \rangle M_T^{-1}\right)  \left( M_T^{-1}\; QQQL + M_T^{-1}\; u^ce^cu^cd^c \right) .
\end{align}

\noindent As a result, compared to the conventional SUSY $SO(10)$-GUT, proton decay rate in our theory is  suppressed by a factor of $\left(M_{\rm PQ}/M_{\rm GUT}\right)^2 \sim \left(10^{11}/10^{16}\right)^2=10^{-10}$.   Although there is no quantitative prediction, this suppression would still allow $p \rightarrow \overline{\nu}K^+$ to be within reach of ongoing and proposed experiments, with a lifetime of order $10^{34}$ yrs., assuming that all SUSY particles that enter into the dressing of the $d=5$ operator have masses of order TeV.\footnote{A similar suppression mechanism utilizing a global $U(1)$  broken around $10^{14}$ GeV has been studied in see Ref. \cite{Du:2014mqa} with the minimal Yukawa sector. However, a suppression factor of $(10^{14}/10^{16})\sim 10^{-2}$ in the amplitude may not be enough to satisfy the experimental lower bound of  the proton lifetime  with TeV scale SUSY particles.} The lack of a definite prediction for the Higgsino mediated proton decay lifetime is understandable,  as it depends on the parameters appearing on the superpotential.   Furthermore, the lifetime depends strongly on the SUSY particle masses, with the rate more suppressed with increasing sfermion masses.  The case of TeV scale SUSY particle masses is of most interest, as it can provide a solution to the gauge hierarchy problem.  We have therefore focused here on the case of TeV scale SUSY spectrum.    

Contrary to the $d=5$ Higgsino mediated proton decay,  the $d=6$ gauge mediated  proton decay rate can be predicted in terms of the GUT scale gauge boson masses.
The dominant decay mode resulting from the $d=6$ baryon number violating operators is is $p\to e^+\pi^0$.
If the threshold corrections arising from the GUT scale are not significant, then the prediction for proton lifetime is similar, but somewhat shorter (due to new contributions from the $X'$ and $Y'$ gauge bosons), compared to SUSY $SU(5)$ prediction. However, the mass of the $(X,Y)$ and $(X',Y')$ gauge bosons depend on the GUT scale threshold effects, which can cause considerable uncertainty in the prediction. The $d=6$ proton lifetime  can be in the range $10^{34-36}$ years, if the GUT scale threshold corrections allow the gauge boson masses to deviate from the GUT scale of $M_{\rm GUT}\sim 2 \times  10^{16}$ GeV by a factor of 3.

We can in fact calculate the Higgsino mediated proton decay rate exactly in our model.  Dressing of the $d=5$ $LLLL$ operators by the Wino gives the most dominant contribution to proton decay, so we shall focus on this operator.  
To compute the decay  rate, we define the relevant amplitude functions $A^{\nu}_{ijk\rho}$ as \cite{Hisano:1992jj,Ellis:1983qm,Babu:1995cw,Babu:1998wi,Babu:2018tfi,Fukuyama:2004xs} as: 
\begin{align}
\scalemath{0.9}{
A^{\nu}_{1bc\rho}= \frac{\lambda^{\prime}_5\langle S_3 \rangle {M^{-1}_T}_{16}{M^{-1}_T}_{61}}{N^2_d} \left( \hat{A}^{\nu}_{1bc\rho}[H^{(1)},H^{(2)}] + x  \hat{A}^{\nu}_{1bc\rho}[F^{(1)},F^{(2)}] + y \hat{A}^{\nu}_{1bc\rho}[H^{(1)},F^{(2)}] + z  \hat{A}^{\nu}_{1bc\rho}[F^{(1)},H^{(2)}] \right),
} \label{amp}
\end{align}

\noindent 
where the amplitude functions  $\hat{A}^{\nu}_{1bc\rho}$ and the Yukawa couplings $H^{(i)}, F^{(i)}$ are  defined in Ref.  \cite{Babu:2018tfi}. The parameters $x,y,z$  are defined as \cite{Goh:2003nv,Babu:2018tfi}:
\begin{align}
x=\left(\frac{\sqrt{3}}{q^{\ast}_2}\right)^2 \frac{{\mathcal{M}^{-1}_T}_{36}{\mathcal{M}^{-1}_T}_{62}}{{\mathcal{M}^{-1}_T}_{16}{\mathcal{M}^{-1}_T}_{61}}, 
\;\;\;
y=\left(\frac{\sqrt{3}}{q^{\ast}_2}\right) \frac{{\mathcal{M}^{-1}_T}_{16}{\mathcal{M}^{-1}_T}_{62}}{{\mathcal{M}^{-1}_T}_{16}{\mathcal{M}^{-1}_T}_{61}},  
\;\;\;
z=\left(\frac{\sqrt{3}}{q^{\ast}_2}\right) \frac{{\mathcal{M}^{-1}_T}_{36}{\mathcal{M}^{-1}_T}_{61}}{{\mathcal{M}^{-1}_T}_{16}{\mathcal{M}^{-1}_T}_{61}}. \label{xyz}
\end{align}

\noindent Here,  $H$ and $F$ are Yukawa coupling matrices obtained within the model from a fit to fermion masses and mixings. The best fit values of these matrices from fermion data are given in Eqs. \eqref{H-value}-\eqref{F-value} in the next section. With these as input one can compute the proton decay rate following the procedure explained in Ref. \cite{Babu:2018tfi}.

\section{Results}

In this section we present our results. First we discuss the fit to the fermion masses and mixings, then we present the details of the proton decay rate calculation, then we discuss the gauge coupling unification within our framework, and finally address lepton flavor violation predicted by the model.

\subsection{Fit to fermion masses and mixings}\label{FIT}
The Yukawa couplings of the model are given in Eq. (\ref{Yuk}).
This is identical to the case of SUSY $SO(10)$ without PQ symmetry.
So we closely follow the definitions and the parametrizations of Ref. \cite{Babu:2018tfi} in the context of non-PQ $SO(10)$ model. Following the same notation as Ref. \cite{Babu:2018tfi}, at the GUT scale, one has the  $SO(10)$ relations: 
\begin{align}
&Y_D= H + F\\
&Y_U= r (H + s F)\\
&Y_E= H -3 F\\
&Y_{\nu_D}= r (H - 3 s F) \label{YNU}\\
&M_N= -\left(v_uY_{\nu_D}\right)^T\left(c_R F\right)^{-1}\left(v_uY_{\nu_D}\right).
\end{align}
Here $Y_{U,D,E}$ are the MSSM Yukawa couplings of up-quarks, down-quarks and charged leptons.  
\noindent We have defined
\begin{align}
& r= \frac{v^{10}_u}{v^{10}_d} \frac{1}{\tan\beta}, \;\;\; s= \frac{v^{126}_u}{v^{126}_d} \frac{v^{10}_d}{v^{10}_u}, \;\;\; c_R= v_R \frac{v_d}{v^{126}_d},  \label{rs-original} \\
&Y_{10}=\frac{v_d}{v^{10}_{d}} H, \;\;\;
 Y_{126}=\frac{v_d}{v^{126}_{d}} F.
\end{align} 

\noindent
Here $v_u$ and $v_d$ are the VEVs of the MSSM fields $H_u^{MSSM}$ and $H_d^{MSSM}$, and we define as usual $\tan\beta = v_u/v_d$.
Although, the light neutrino masses receive contributions from type-I seesaw as well as type-II seesaw, the magnitude of the type-II contribution comes out to be small, because the weak triplets in the model have masses of order the GUT scale. Thus we work in the type-I seesaw dominance scenario \cite{ss1,ss2,ss3,ss4}.

The minimal Yukawa sector of   SUSY $SO(10)$-GUT has 12 real parameters and 7 phases to fit 18 observables. 
To fit the fermion masses and mixings we perform a $\chi^2$-analysis   at the GUT scale, which we take to be
$M_{\rm GUT}= 2\times 10^{16}$ GeV. We minimize the function  $\chi^2= \sum_i  P^2_i$  to achieve the best fit, where the pull is defined as  $P_i=(O_{i\;th}-E_{i\;exp})/ \sigma_i$. 
Here $\sigma_i$ represents experimental $1\sigma$ uncertainty and $O_{i\;th}$ and  $E_{i\;exp}$ represent  theoretical prediction and experimental central value of observable
$i$. To get the GUT scale values
of the observables in the charged fermion sector we take the central values at the $M_Z$
scale from Table-1 of Ref. \cite{Antusch:2013jca}. For neutrino observables, the low energy values are taken from Ref. \cite{deSalas:2017kay}.   With these inputs, we do the RGE running of the Yukawa couplings \cite{Machacek:1983fi,Arason:1991ic}, the CKM parameters \cite{Babu:1987im} and the effective couplings of the neutrino $d=5$ operator, $\kappa$ \cite{Babu:1993qv,Chankowski:1993tx,Antusch:2001ck} within the SM up to the SUSY-scale, which we choose  to be  1 TeV. Above 1 TeV the full MSSM is restored, so we use the  relevant MSSM RGEs \cite{Barger:1992ac, Barger:1992pk, Antusch:2001vn} and evolve them up to the GUT scale. For the charged fermion observables, we fit to these evolved values at the GUT scale. Since the requirement from the neutrino data is to have right-handed neutrinos at some lower scale than the GUT scale in the type-I seesaw scenario, in the fit procedure we include the  threshold corrections due to the right-handed neutrinos from the $v_R$ scale to the GUT scale \cite{Babu:2018tfi} (see also Ref. \cite{Fukuyama:2016mqb}). So above the SUSY scale,  the effective couplings of the neutrino $d=5$ operator running is performed up to the intermediate scale instead of the GUT scale. 

For our numerical  analysis we fix $\tan\beta=10$. 
Our best fit observables are presented in Table \ref{result}.  We see from this Table that the model gives an excellent fit to the data, with a total $\chi^2 = 6.3$.  Most of the contributions to $\chi^2$ arises from the $d$-quark mass, which is 2.3 $\sigma$ below the central value.  All other observables are within 1 $\sigma$, providing an excellent overall fit.

In Table \ref{pred} we list the predictions of the model in the fermion sector for quantities that are currently unknown.  The central value of the CP violating Dirac phase relevant for neutrino oscillations is found to be $\delta^{PMNS} = 327^0$.  However, it should be noted that 
large deviations from the best fit are possible, owing to existence of nearby local minima with acceptable $\chi^2$.  
The possible variation of this phase as a function of $\chi^2$ was presented in Fig. 1 of  Ref.  \cite{Babu:2018tfi}. Similar results should also hold in our current setup, although we have not investigated this in detail.

In the neutrino sector, the ordering of masses is normal, with the effective mass for neutrinoless double beta decay found to be $m_{\beta \beta} \simeq 5$ meV.  While this is well below the current limit from Kamland-Zen experiment of $(61-165)$ meV \cite{kamlandzen}, future improvements can potentially probe this prediction.  

The Yukawa coupling  parameters corresponding to the best fit are given by:
\begin{align}\label{rs-values}
 r=9.43514,\;s= 0.327008 - 0.0387752 i, \;c_R= 2.90638\times 10^{14} \;\rm{GeV}.
\end{align}
\begin{align}\label{F-value}
F=
\left(
\scalemath{0.85}{
\begin{array}{ccc}
4.81752\times 10^{-5}  & 0 & 0 \\
 0 & 2.02566\times 10^{-3}  & 0 \\
 0 & 0 & 7.04398\times 10^{-3} 
\end{array}
}
\right),
\end{align}
\begin{align}\label{H-value}
H=
\left(
\scalemath{0.82}{
\begin{array}{ccc}
1.45607\times 10^{-4} \, -2.45964\times 10^{-4}  i & -1.52894\times 10^{-3} +6.83905\times 10^{-4}  i
   & -2.74948\times 10^{-3} -2.48324\times 10^{-3}  i \\
 -1.52894\times 10^{-3} +6.83905\times 10^{-4}  i & 8.63688\times 10^{-3} \, +1.50766\times 10^{-3}  i &
   5.35959\times 10^{-3} \, +2.03718\times 10^{-2}  i \\
 -2.74948\times 10^{-3} -2.48324\times 10^{-3}  i & 5.35959\times 10^{-3} \, +2.03718\times 10^{-2}  i &
   -3.84776\times 10^{-2} +2.96883\times 10^{-2}  i
\end{array}
}
\right).
\end{align}

\FloatBarrier
\begin{table}[t!]
\centering
\footnotesize
\resizebox{1.1\textwidth}{!}{
\begin{tabular}{|c|c|c||c|c|c|}
\hline
\pbox{10cm}{Masses \\(in GeV)} & \pbox{10cm}{~~~~~Inputs \\ ($\mu= M_{\rm GUT}$)} & \pbox{25cm}{Fitted values (pulls) \\ ($\mu= M_{\rm GUT}$)}  & \pbox{10cm}{Mixing parameters and\\Mass squared differences} & \pbox{10cm}{~~~~~Inputs \\ ($\mu= M_{\rm GUT}$)} & \pbox{25cm}{Fitted values (pulls) \\ ($\mu= M_{\rm GUT}$)}    \\ [1ex] \hline

$m_{u}/10^{-3}$ & 0.502$\pm$0.155 & 0.518 (0.103)&$|V_{us}|/10^{-2}$  & 22.54$\pm$0.07 & 22.54 (0.025)  \\ \hline
$m_{c}$   & 0.245$\pm$0.007 & 0.246 (0.131)&$|V_{cb}|/10^{-2}$  & 3.93$\pm$0.06 & 3.95 (0.405)  \\ \hline
$m_{t}$   & 90.28$\pm$0.89 & 90.29 (0.007)&$|V_{ub}|/10^{-2}$  & 0.341$\pm$0.012 & 0.340 (-0.017) \\ \hline

$m_{d}/10^{-3}$  & 0.839$\pm$0.17 & 0.451 (-2.31)&$\delta_{CKM}$ & 1.208$\pm$0.054  & 1.227 (0.360)  \\ \hline
$m_{s}/10^{-3}$  & 16.62$\pm$0.90 & 17.02 (0.449)&$\Delta m^{2}_{sol}/10^{-5}$(eV$^{2}$) & 9.039$\pm$0.227 & 9.012 (-0.117)   \\ \hline
$m_{b}$  & 0.938$\pm$0.009 & 0.934 (-0.507)&$\Delta m^{2}_{atm}/10^{-3}$(eV$^{2}$) & 3.051$\pm$0.051 & 3.057 (- 0.120)  \\ \hline

$m_{e}/10^{-3}$   & 0.344021 & 0.344051 (0.089)&$\sin^{2}\theta^{\rm{PMNS}}_{12}$ & 0.3219$\pm$0.017 & 0.3177 (-0.188)  \\ \hline
$m_{\mu}/10^{-3}$  & 72.6256 & 72.6262 (0.008)&$\sin^{2}\theta^{\rm{PMNS}}_{23}$ & 0.431$\pm$0.019 & 0.4368 (-0.310)  \\ \hline
$m_{\tau}$  & 1.24038 & 1.24045 (0.061)&$\sin^{2}\theta^{\rm{PMNS}}_{13}$ & 0.0216$\pm$0.00082 & 0.02166 (0.081)   \\ \hline

\end{tabular}
}
\caption{ Best fit values of the observables corresponding to the  type-I dominance seesaw scenarios for $\tan\beta= 10$. This best fit corresponds to total $\chi^2=6.3$. For the associated $1\;\sigma$ uncertainties of the observables at the GUT scale, we keep the same percentage uncertainty with respect to the central value of each quantity as that at the $M_Z$ scale. For the charged lepton Yukawa couplings at the GUT scale, a relative uncertainty of $0.1\%$ is assumed in order to take into account the theoretical uncertainties arising for example from threshold effects. }
\label{result}
\end{table}

\begin{table}[th!]
\centering
\footnotesize
\resizebox{0.7\textwidth}{!}{
\begin{tabular}{|c|c|}
\hline
Quantity & \pbox{10cm}{Predicted Value }  \\ [1ex] \hline
$\{m_{1}, m_{2}, m_{3} \}$ (in eV) & $\{ 3.57\times 10^{-3}, 1.01\times 10^{-2}, 5.54\times 10^{-2} \}$   \\ \hline

$\{\delta^{PMNS}, \alpha^{PMNS}_{21}, \alpha^{PMNS}_{31} \}$ & $\{ 327.46^{\circ}, 29.35^{\circ}, 65.6^{\circ} \}$  \\ \hline

$\{m_{cos}, m_{\beta}, m_{\beta \beta} \}$ (in eV) & $\{ 6.91\times 10^{-2}, 6.73\times 10^{-3}, 4.99\times 10^{-3} \}$  \\ \hline

$\{M_{1}, M_{2}, M_{3} \}$ (in GeV)  & $\{ 1.40\times 10^{10}, 5.88\times 10^{11}, 2.04\times 10^{12} \}$ \\ \hline
\end{tabular}
}
\caption{ Predictions corresponding to the best fit values presented in Table \ref{result} for type-I dominance seesaw scenario.  $m_{i}$ are the light neutrino masses, $M_{i}$ are the right handed neutrino masses, $\alpha_{21,31}$ are the Majorana phases following the PDG parametrization, $m_{cos}=\sum_{i} m_{i}$, $m_{\beta}=\sum_{i} |U_{e i}|^{2} m_{i}$ is the effective mass parameter for beta-decay and $m_{\beta \beta}= | \sum_{i} U_{e i}^{2} m_{i} |$ is the effective mass parameter for neutrinoless double beta decay.}
\label{pred}
\end{table}

\newpage
\subsection{Proton decay calculation}
For the proton decay rate calculation, we proceed in the following way:

\begin{enumerate}

\item[--] We first solve the stationary conditions corresponding to the superpotential of Eq. \eqref{potential}.  These conditions, which can be found in  Ref. \cite{Fukuyama:2004ps, Babu:2016cri},  are solved for the mass parameters $E, \Phi_1, m_1, m_2$ and $m_5$. Note that $m_5$ does not play any role for proton decay calculation. We follow the  the symmetry breaking pattern shown in Eq. \eqref{chain}.  
 
\item[--] We set $\lambda_1=1$ and $\Phi_3=M_{\rm GUT}$. We choose $M_{\rm GUT}= 2\times 10^{16}$ GeV. Furthermore, we fix $v_R= \overline{v}_R = \langle S_3\rangle = 10^{12}$ GeV. Then the dimensionless free parameters of the theory are: $\lambda_{2}, \lambda_{3}, \lambda_{4}^{\prime}, \lambda_{5}^{\prime}, \lambda_{10}$ and  $\lambda_{13}$  which we take to be real. There is one free parameter with dimension of mass $\Phi_2$, which is taken to be complex. We scan over this parameter set $\{ \lambda_{2}, \lambda_{3}, \lambda_{4}^{\prime}, \lambda_{5}^{\prime}, \lambda_{10}, \lambda_{13}, \Phi_2 \}$ such that the dimensionless couplings are restricted to be $|\lambda_i^{(\prime)}|\leq 2$ for perturbativity.  

\item[--]  On the doublet mass matrix Eq. \eqref{doublet}, we impose the determinant $=0$ condition to get the MSSM Higgs pair at the EW-scale. A mini fine-tuning condition is further imposed  to keep another pair of Higgs doublet at the PQ scale for the consistency of the theory as discussed earlier. These fine-tuning conditions are imposed in such  a way that the masses of the rest of the Higgs doublets and all the color-triplets are kept around the GUT scale. 

\item[--]  The numerical values of  $r$ and $s$ parameters defined in Eq. \eqref{rs} are fixed by the best fit of the fermion spectrum given in Eq. (\ref{rs-values}). While computing the proton decay rate, we also impose the constraints that the superpotential parameters correctly reproduce   $r$ and $s$ values given in Eq. \eqref{rs-values}.  

\item[--] With each parameter set, satisfying all of the above constraints, we compute the    proton decay    amplitude function defined in Eq. \eqref{amp} that   contains the Yukawa couplings, for which we use the best fit values given in Eqs. \eqref{F-value}-\eqref{H-value} by going to the physical basis of fermions (for details see Ref. \cite{Babu:2018tfi}).  

\item[--] By using the amplitude function Eq. \eqref{amp} calculated as above, we estimate the minimum value of the sfermion mass, $m_S$  that satisfies all the experimental  proton decay bounds: $\tau_p(p\to K^+\overline{\nu})\geq 5.9\times 10^{33}$ years and  $\tau_p(p\to \pi^+\overline{\nu})\geq 3.9\times 10^{32}$ years \cite{Hayato:1999az}. For this computation we fix the Wino mass to be $m_W=$ 1 TeV. We have taken the relevant nuclear matrix elements  from Ref.  \cite{Aoki:2017puj}.

\end{enumerate}

\begin{figure}[t!]
\begin{center}
\includegraphics[width=16.cm]{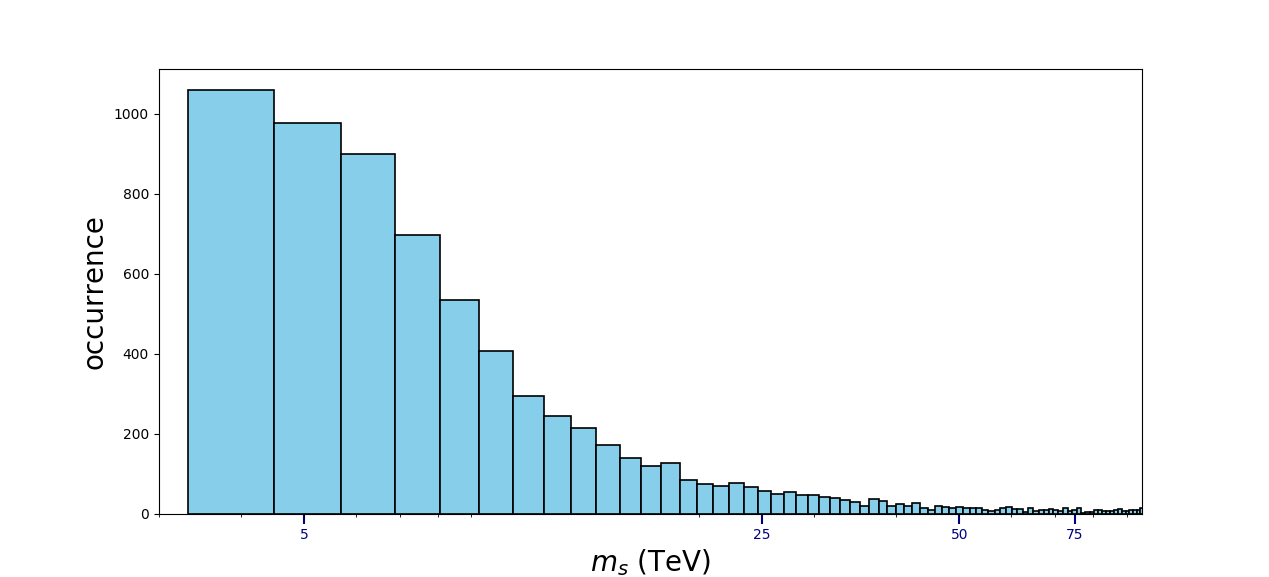}
\end{center}
\caption{ Histogram showing the probability distribution of the minimum required sfermion mass, $m_S$ that satisfies  the proton lifetime bounds. We allow the range 3 TeV $\leq m_S \leq$ 100 TeV with fixed Wino mass, $m_W=$ 1 TeV.  }
\label{histogram}
\end{figure} 

In Fig. \ref{histogram}, we present our result on the allowed range for SUSY scalar masses arising from proton lifetime limits obtained by following the procedure outlined above. This figure shows the  probability distribution as a function of the minimum required sfermion mass, $m_S$, to satisfy the experimental lower bounds on proton lifetime when $m_S$ is varied in the range 3 TeV $\leq m_S \leq$ 100 TeV.  We see that, owing to the suppression in the rate from PQ symmetry, SUSY scalar masses of 3 TeV is fully consistent.  Even lower value, e.g., 1 TeV is found to be consistent with proton decay. 
In the analysis performed above, for simplicity, we have not taken into account the threshold corrections
  of the particles live around the GUT scale. Incorporating the threshold corrections may broaden the parameter space presented in this work.  

A sample point  that is consistent with the proton decay bounds for $m_S=$ 3 TeV is given by the following parameter set:
\begin{align}
&\lambda_{2}=-1.614315, \;
\lambda_{3}=-0.371666, \;
\lambda_{4}^{\prime}=0.005410,\; 
\lambda_{5}^{\prime}=-1.290149, 
\nonumber \\ &
\lambda_{10}=0.327337, \;
\lambda_{13}=0.749386, \;
\frac{\Phi_2}{M_{\rm GUT}}= -0.500789-0.0205743\;i .
\nonumber
\end{align}   

\noindent
With this parameter set, the stationary conditions fix:
\begin{align}
&\frac{V_E}{M_{\rm GUT}}=-2.503066+0.079884\;i, \;
\frac{\Phi_1}{M_{\rm GUT}}=0.433696+0.017817\;i ,
\nonumber \\ &
\frac{m_1}{M_{\rm GUT}}=0.16479-0.000951\;i, \;
\frac{m_2}{M_{\rm GUT}}=0.132849-0.001174\;i.
\nonumber
\end{align}

\noindent
And the corresponding values of the $r$ and $s$ parameters are given by: 
\begin{align}
r= 9.43514,\;\;
s= 0.32700-0.03877\;i .
\nonumber
\end{align}

\noindent
Note that, for the above sample point, $\lambda_5^{\prime}\langle S_3 \rangle \sim 10^{12}$ GeV. Below, we  present a second consistent sample point for which $\lambda_5^{\prime}\langle S_3 \rangle \sim 10^{11}$ GeV. This second sample point  that is consistent with the proton decay bounds for $m_S=$ 3 TeV and reproduces $r$ and $s$ exactly is given by the following parameter set:
\begin{align}
&\lambda_{2}=1.77088, \;
\lambda_{3}=-0.022657, \;
\lambda_{4}^{\prime}=-0.017888,\; 
\lambda_{5}^{\prime}=-0.241662, 
\nonumber \\ &
\lambda_{10}=-1.368922, \;
\lambda_{13}=1.391076, \;
\frac{\Phi_2}{M_{\rm GUT}}= -0.500789-0.0205743\;i .
\nonumber
\end{align}   

We see from the analysis above that proton lifetime for decay into $\overline{\nu} K^+$ is close to the current experimental limit.  While it is difficult to make this statement more precise owing to uncertainties in the superpotential parameters, we expect proton decay to be within reach of ongoing and forthcoming deep underground experiments.

\subsection{Gauge coupling unification}

\begin{table}[b!]
\centering
\footnotesize
\resizebox{1\textwidth}{!}{
\begin{tabular}{|c|c|c|c|c|c|}
\hline
Multiplet, $\phi$&\pbox{10cm}{Running coefficient\\ ($b_1, b_2, b_3$) }& $M_{\phi}$ & $M_{\rm GUT}$ & $\alpha^{-1}_{\rm GUT}$ & $\tau_p(p\to e^+\pi^0)$ in yrs  \\ [1ex] \hline
$(6,1,\frac{1}{3})+c.c.$&$(\frac{4}{5},0,5)$& $2.76\times 10^{15}$ GeV &$1.65\times 10^{16}$&$24.3$ &$7.39\times 10^{35}$  \\ \hline
$(6,1,-\frac{2}{3})+c.c.$&$(\frac{16}{5},0,5)$& $2.76\times 10^{15}$ GeV  &$9.92\times 10^{15}$&$24.46$  &$9.78\times 10^{34}$\\ \hline
$(6,1,\frac{4}{3})+c.c.$&$(\frac{64}{5},0,5)$& $2.75\times 10^{15}$ GeV &$5.0\times 10^{15}$&$24.68$  &$6.42\times 10^{33}$\\ \hline
\end{tabular}
}
\caption{ Three example scenarios of gauge coupling unification which invoke small threshold correction from a single color-sextet multiplet and its conjugate and the corresponding $d=6$ proton decay lifetime. 
Concerning the quoted values of the proton decay lifetime,
we stress the fact that
 these are just  estimates, since we have not included the threshold corrections from the other fields except the color-sextet.  Each of these scenarios can be considered viable once all the threshold corrections are taken into account.
}
\label{table-coupling}
\end{table}

\begin{figure}[th!]
\begin{minipage}{0.1\textwidth}
(a) 
\end{minipage}
\begin{minipage}{0.7\textwidth}
\begin{center}
\includegraphics[width=9.5cm]{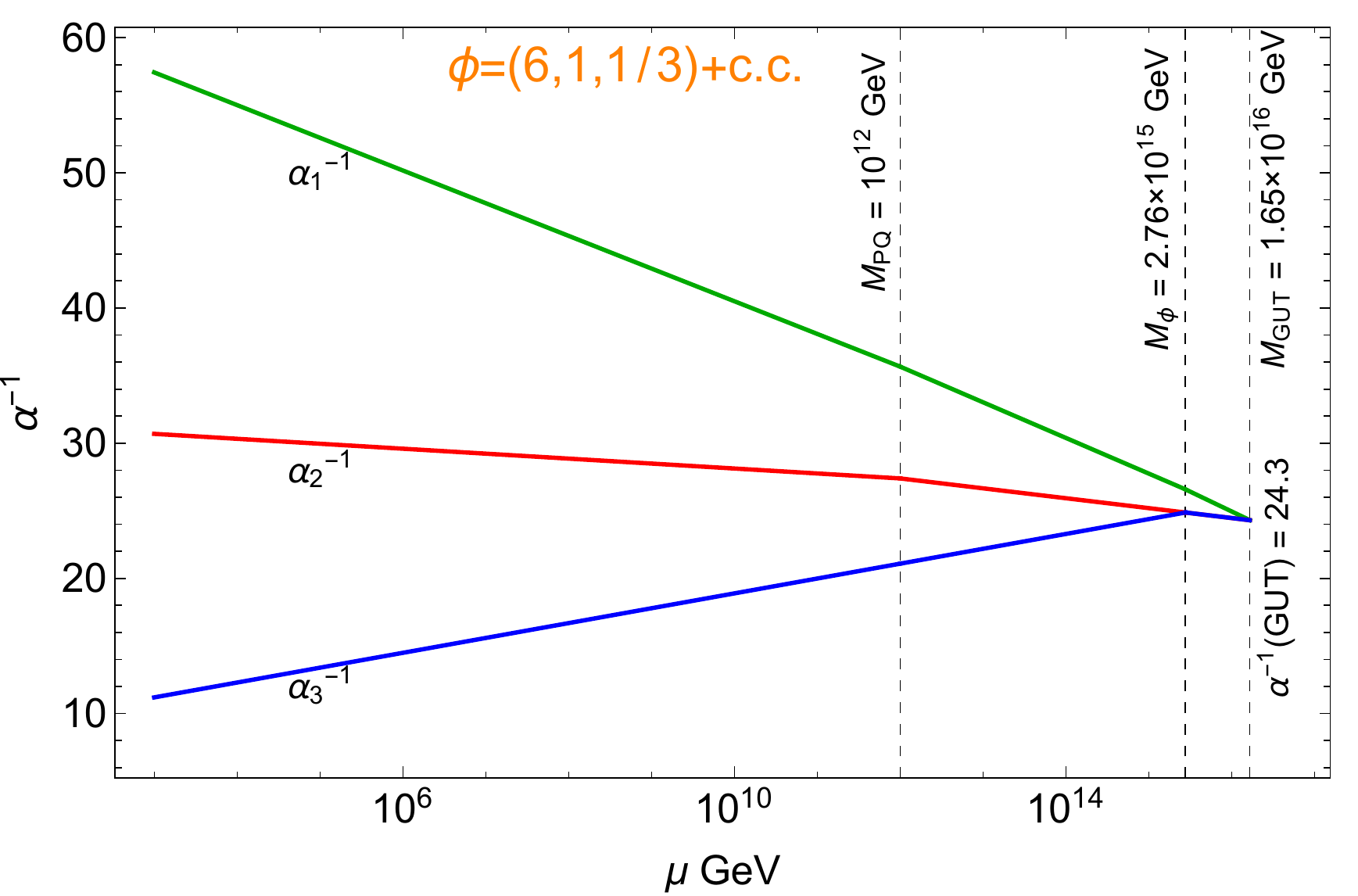}
\end{center}
\end{minipage}\\
\begin{minipage}{0.1\textwidth}
(b) 
\end{minipage}
\begin{minipage}{0.7\textwidth}
\begin{center}
\includegraphics[width=9.5cm]{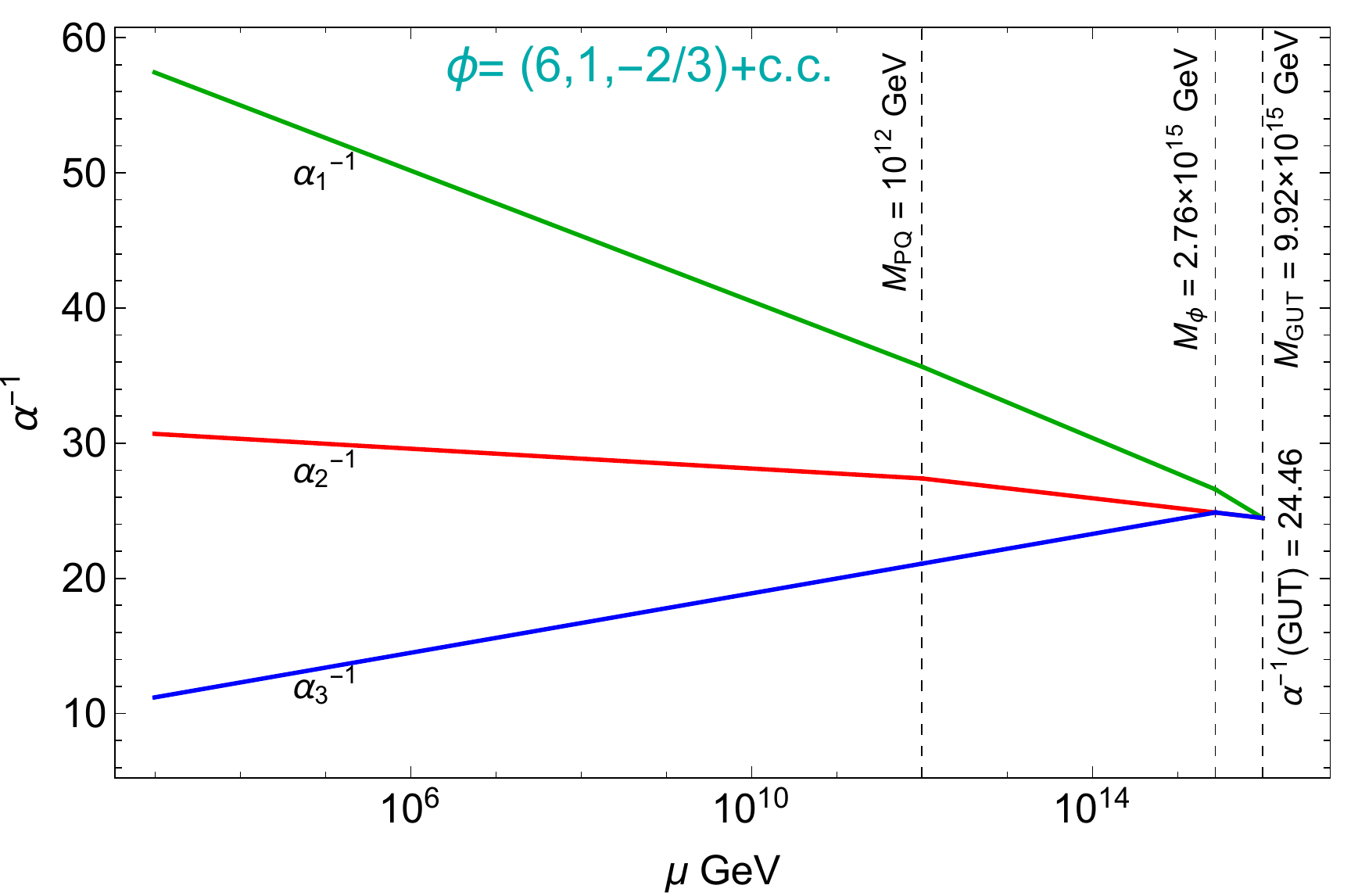}
\end{center}
\end{minipage}\\
\begin{minipage}{0.1\textwidth}
(c) 
\end{minipage}
\begin{minipage}{0.7\textwidth}
\begin{center}
\includegraphics[width=9.5cm]{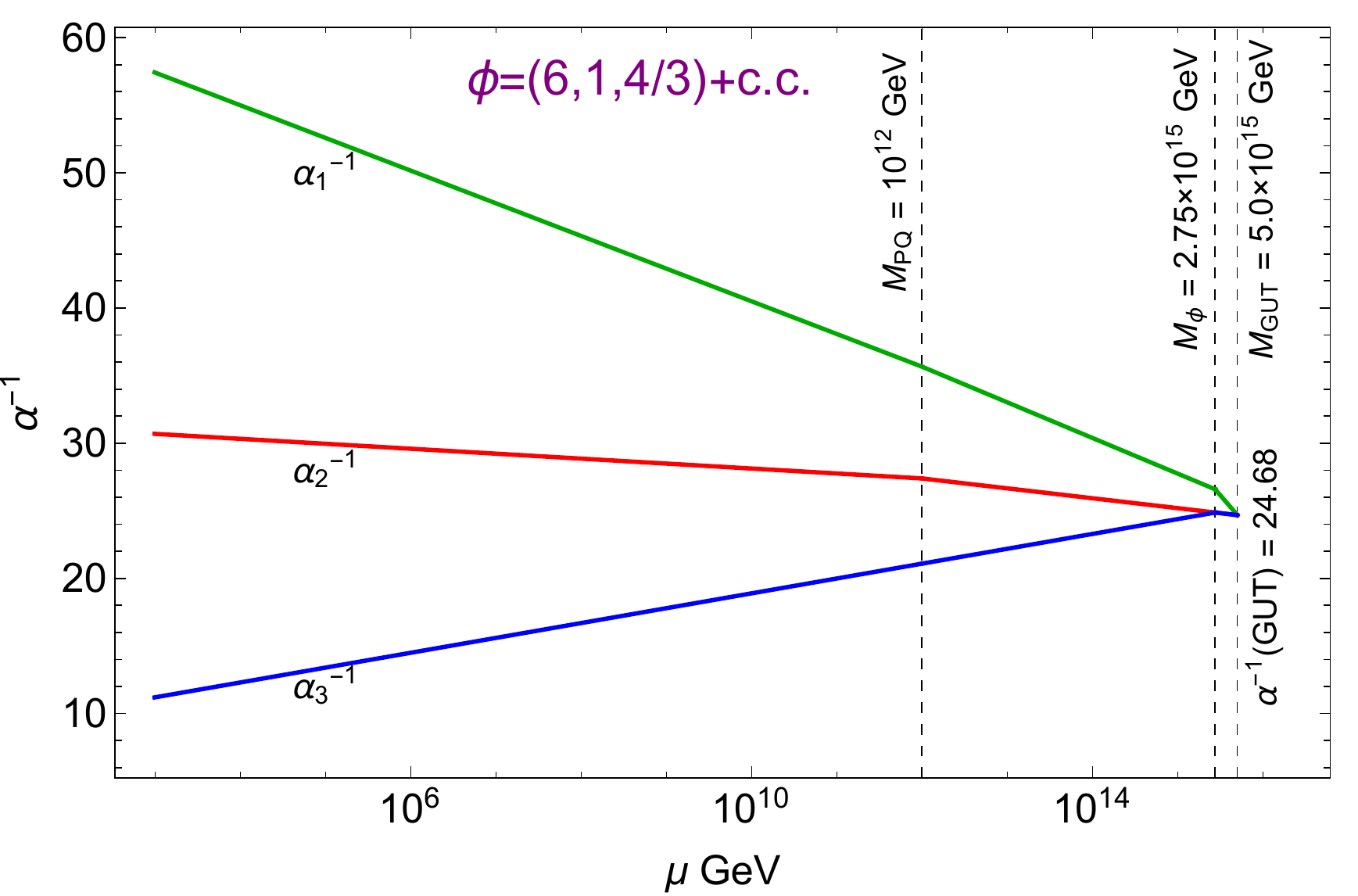}
\end{center}
\end{minipage}
\caption{ Evolution of gauge couplings including effects of a second pair of Higgs doublets at $M_{\rm PQ}= 10^{12}$ GeV. A color sextet field ($\phi$) and its conjugate is assumed to have a mass slightly below the GUT scale, with the mass scale chosen to restore unification of gauge couplings. The three plots correspond to the three color sextet choices (a) $(6,1,1/3)$, (b) $(6,1,-2/3)$ and (c) $(6,1,4/3)$ with each threshold effect listed in Table \ref{table-coupling}. For this illustration, we set $M_{\rm SUSY}=1$ TeV and used the one-loop RG evolution of gauge couplings.   }
\label{unification}
\end{figure}

We now comment on the gauge coupling unification in our model. Note that the automatic   gauge coupling unification of the MSSM is not realized in our scenario. This is because of the appearance of an additional $SU(2)_L$ doublet Higgs fields at the PQ breaking scale.
However, since the doublet contributions to the running of gauge couplings are not very significant, small threshold corrections near  the GUT scale can restore the coupling unification. With the MSSM spectrum all the way to the PQ symmetry scale, and an extra pair of doublets at the PQ scale, unification can be realized by lowering the mass of a single GUT-scale multiplet carrying color. It is sufficient to lower the mass by an order of magnitude below the GUT scale.
To demonstrate gauge coupling unification within our set-up, we stick to this simple scenario where only one pair of fields has a mass around $10^{15}$ GeV.  We choose this multiplets to be a color-sextet, isospin-singlet field and its conjugate. In our model three such fields are present: $(6,1,1/3)$, $(6,1,-2/3)$ and $(6,1,4/3)$, along with their conjugate fields.  The possibilities  of coupling unification with these color-sextets are summarized in Table. \ref{table-coupling} and the corresponding unification plots are presented in Fig. \ref{unification}.  To obtain this results, we first run the 1-loop MSSM gauge coupling constants from 1 TeV to the PQ scale that we fix to be $10^{12}$ GeV, where a pair of Higgs doublet lies; then from the PQ scale to the  color-sextet mass scale including the RGE effects of the extra doublet pair; and from there upto the unification scale including the effects of the color sextet fields. The color-sextet mass  is appropriately chosen to achieve unification. At the TeV  scale the gauge couplings are taken to be  $g_1=0.46774, g_2=0.63990$ and $g_3=1.0597$ \cite{Antusch:2013jca}.    Among these three different options,  lowering the mass of the $(6,1,-2/3)$-multiplet is the most favorable. The mass of this multiplet is given by  $(m_5-2\sqrt{3/5} \lambda_8 E)$, which  can be tuned to a value below the GUT scale without making other multiplets light.  Also, in this mass formula,    the dimensionless parameter $\lambda_8$  is free and does not play any role in the proton decay calculation. (The mass parameter $m_5$ also does not enter in proton decay calculation and  is fixed by one of the stationary conditions). The corresponding unification scale is found to be $\sim 10^{16}$ GeV and the unified gauge coupling is $\alpha^{-1}_{\rm GUT}=24.46$, with these, the expected $d=6$ proton decay $p\to e^+\pi^0$ lifetime   can be estimated \cite{Babu:2010ej}  to be $\tau_p\sim 9\times 10^{34}$ yrs, which is above the current experimental lower limit set by Super-Kamiokande, $\tau_p > 1.6\times 10^{34}$ yrs \cite{sk2}. In Table. \ref{table-coupling}, along with the unification constraints, we also estimate the  lifetime for the other two cases. For the case of $(6,1,4/3)$, the lifetime for the   $p\to e^+\pi^0$ mode came out to be smaller than the present experimental  lower limit. However, these are just  estimates, since we have not included the threshold corrections from the other fields except the color-sextet.  Each of these scenarios can be considered viable once all the threshold corrections are taken into account.


\subsection{Lepton Flavor Violation}

In this section we discuss  lepton flavor violating (LFV) rare decays predicted by our model.  Assuming flavor universality of the SUSY breaking mass parameters at the GUT scale, the main source of LFV in SUSY $SO(10)$ is the presence of the Dirac neutrino coupling $Y_{\nu}\ell \nu^c H_u$ of Eq. \eqref{YNU}, which is a combination of Yukawa coupling of the $10_H$ and $\overline{126}_H$ multiplets. These $\nu^c$ fields have masses of order $v_R \sim 10^{12}$ GeV, and the renormalization group running of the SUSY mass parameters in the momentum range $v_R \leq \mu \leq M_{\rm GUT}$ would impart the flavor structure of the neutrino Dirac Yukawa coupling to the sfermions.  
This RGE-induced off-diagonal entries of the left-handed slepton mass matrix in the leading log approximation can be expressed as:
\begin{align}
(\Delta^{\ell}_{LL})_{ij}=-\frac{3m_0^2+A^2_0}{8\pi^2}\sum^3_{k=3} (Y_{\nu})_{ik} (Y^{\ast}_{\nu})_{kj} \text{ln}(\frac{M_{\rm GUT}}{M_{R_k}}),
\end{align}  

\noindent
where, $Y_{\nu}$ is the Dirac neutrino Yukawa coupling in a basis where the charged lepton and right-handed neutrino mass matrices are diagonal.
Assuming scalar mass universality and gaugino mass unification as is usually done in constrained MSSSM, the number of parameters in the SUSY breaking sector is reduced to the set: $\{m_0, m_{1/2}, A_0, \text{sgn}(\mu), \tan\beta \}$, where, $m_0$ is the common scalar mass, $m_{1/2}$ is the common gaugino mass, $A_0$ the universal tri-linear coupling and $\mu$ the Higgs mass term. For LFV calculation, we restrict ourselves to the cMSSM scenario that will also fix the $\mu$ parameter from electroweak symmetry breaking condition.  We choose sgn($\mu) >0$ (although the LFV results are essentially the same for negative $\mu$), and set $\tan\beta=10$ corresponding to the fermion fit. The fit to fermion spectrum fully determines the Dirac coupling matrix $Y_{\nu}$ and the right-handed neutrino masses $M_{R_k}$ in our framework. As a result we can now compute the rates for LFV  processes as  functions of only the SUSY breaking parameters.  With our choice, this set is reduced to $\{m_0, m_{1/2}, A_0\}$. 

As is well recognized, cMSSM with TeV scale scalar masses cannot reproduce a Higgs mass of 125 GeV, unless $A_0$ takes rather large values of order 10 TeV.  Here we are interested in LFV processes within MSSM that can correctly produce the Higgs mass as well. We shall thus allow for some new contributions, for example from stop quarks which have non-universal masses, for the Higgs mass.  
For our numerical analysis, we will fix few different values of $A_0$ of order TeV.

For the calculation of LFV,  first we write down the relevant MSSM sparticle masses at the $M_Z$ scale. The gaugino, sleptons and squark  masses are given by \cite{Babu:2008ge}:
\begin{align}
&
\{M_1, M_2, M_3\}= \{ 0.412, 0.822, 2.844 \} m_{1/2}, \\
&
\widetilde{m}^2_{Q_{1,2}}=6.79m^2_{1/2}+m^2_0,\;\;
\widetilde{m}^2_{U_{1,2}}=6.37m^2_{1/2}+m^2_0,\;\;
\widetilde{m}^2_{D_{1,2}}=6.32m^2_{1/2}+m^2_0,\\
&
\widetilde{m}^2_{L_{1,2,3}}=0.52m^2_{1/2}+m^2_0,\;\;
\widetilde{m}^2_{E_{1,2,3}}=0.15m^2_{1/2}+m^2_0.
\end{align} 

\noindent
Furthermore, the $\mu$ term is fixed from the symmetry breaking constraint:
\begin{align}
M^2_Z=-2.0\mu^2+5.44m^2_{1/2}+0.183m^2_0+0.2A^2_0-0.87m_{1/2}A_0.
\end{align}

\noindent
Now, the branching ratio of $\ell_i \to \ell_j \gamma$ is given by:
\begin{align}
BR(\ell_i \to \ell_j \gamma)=\frac{48\pi^3\alpha}{G^2_F}\left( |A^{ij}_L|^2 + |A^{ij}_R|^2 \right) .
\end{align}

\begin{figure}[th!]
\begin{minipage}{0.1\textwidth}
(a) 
\end{minipage}
\begin{minipage}{0.7\textwidth}
\begin{center}
\includegraphics[width=9.5cm]{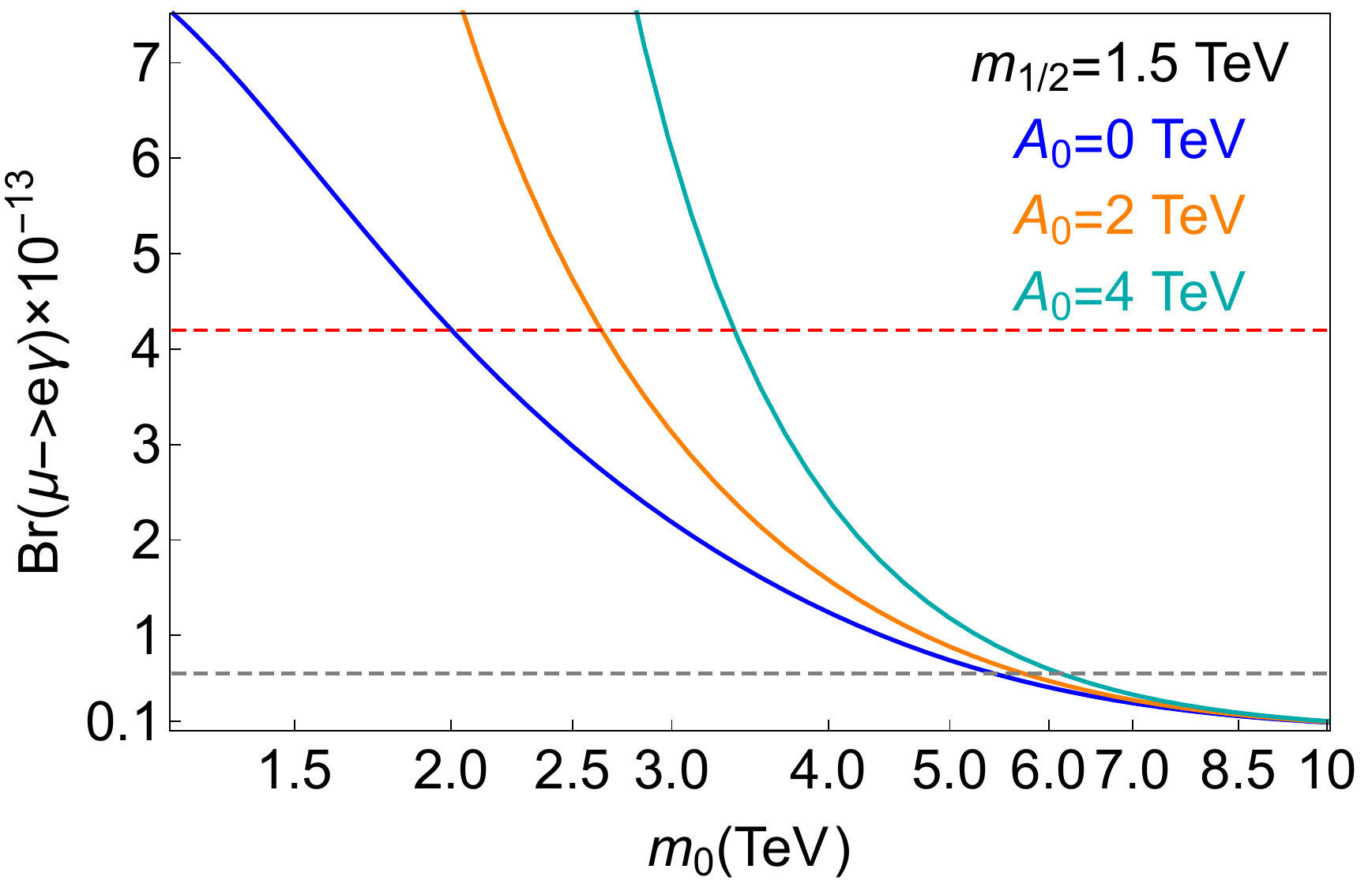}
\end{center}
\end{minipage}\\
\begin{minipage}{0.1\textwidth}
(b) 
\end{minipage}
\begin{minipage}{0.7\textwidth}
\begin{center}
\includegraphics[width=9.5cm]{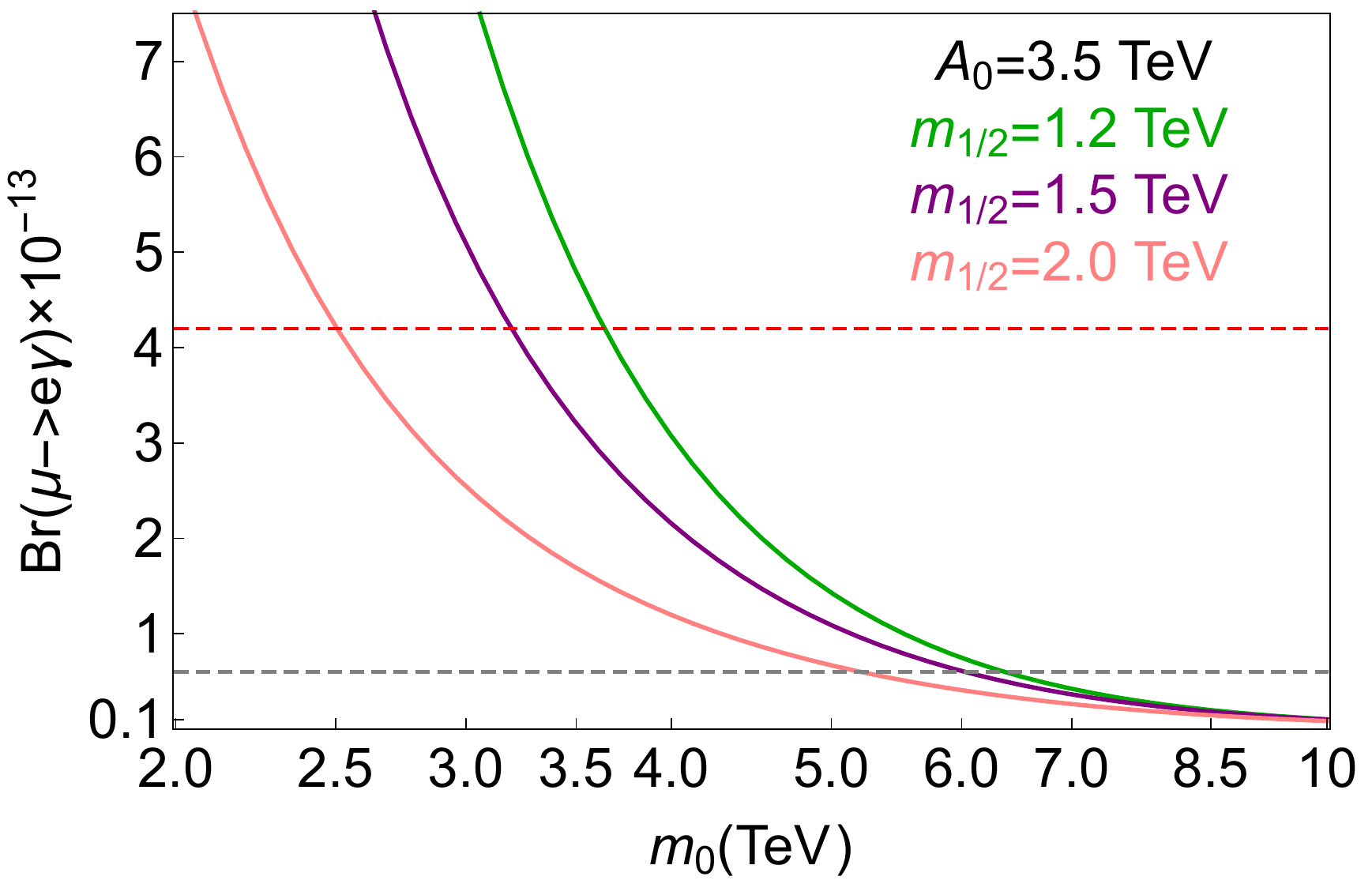}
\end{center}
\end{minipage}
\caption{ The branching ratio of $\mu \to e\gamma$ is presented for our SUSY $SO(10)$ framework. We set $\rm sgn(\mu) >0$ and $\tan\beta =10$.  In the Fig. (a) we fix $m_{1/2} = 1.5$ TeV and show variations for three values of $A_0$.  In the Fig. (b)  we fix $A_0 = 3.5$ TeV and show variations in three values of $m_{1/2}$.  The red horizontal dashed line corresponds to the current  upper bound $Br(\mu \to e\gamma) < 4.2\times 10^{-13}$  by the MEG experiment  and the  gray horizontal dashed line represents the projected sensitivity $Br(\mu \to e\gamma) < 6\times 10^{-14}$ by the    MEG II experiment  \cite{TheMEG:2016wtm}.  }
\label{mu-e}
\end{figure} 

\noindent
The amplitude functions $A_{L,R}$  can be found in Refs.  \cite{Hisano:1995cp,Paradisi:2005fk,Ciuchini:2007ha}. The amplitudes are evaluated in the mass insertion  approximation and we include loop contributions from both the neutralinos and charginos.   For our model, the relevant amplitude is $A_L= A_{L2}+A_{L1}$ with, $A^{ij}_{L2}\propto (\Delta^{\ell}_{LL})_{ij}$ and $A^{ij}_{L1}\propto (\Delta^{\ell}_{RL})_{ij}$, where, $(\Delta^{\ell}_{RL})_{ji}= m_i/\widetilde{m}_L \tan\beta  (\Delta^{\ell}_{LL})_{ji}$. From our fermion fit presented in Sec. \ref{FIT}, we compute: 
\begin{align}
\Delta^{\ell}_{LL}= (3m_0^2+A^2_0)\times  10^{-3}
\left(
\begin{array}{ccc}
 0 & 0.6821\, -0.4617 i & -0.5025-2.136 i \\
 0.6821\, +0.4617 i & 0 & -4.165+11.121 i \\
 -0.5025+2.136 i & -4.165-11.121 i & 0 \\
\end{array}
\right).
\end{align}

\noindent
Using these as inputs, we compute the lepton flavor violating rare decay $\mu \to e\gamma$. Our results are presented in Fig. \ref{mu-e}. In this figure, the red horizontal dashed line corresponds to the current  upper bound $Br(\mu \to e\gamma) < 4.2\times 10^{-13}$  by the MEG experiment  and the  gray horizontal dashed line represents the projected sensitivity  $Br(\mu \to e\gamma) < 6\times 10^{-14}$ by the    MEG II experiment  \cite{TheMEG:2016wtm}.   The  Fig. \ref{mu-e} (a) shows variations of the branching ratio as a function of $m_0$ for different values of $A_0$ and a fixed value of $m_{1/2} = 1.5$ TeV.  The  Fig. \ref{mu-e} (b) shows the same variation for different values of $m_{1/2}$ and $A_0$ fixed at 3.5 TeV.  From these figures we see that part of the parameter space is excluded by the current experimental limit, and that future improvement can probe SUSY spectrum as large as 10 TeV.

We note that the branching ratio for the decay $\tau \to \mu \gamma$ is suppressed in our framework.  When we choose parameters to satisfy the current limit on $\mu \to e\gamma$ branching ratio, we obtain
$Br(\tau \to \mu \gamma)=8\times 10^{-11}$, which is well below the experimental limit. The rare decay processes $\ell_i\to \ell_j\gamma$ provides the most stringent constraints on the parameter space 
of this class of models.  Other LFV processes such as 
$\ell_i \to 3\ell_j$ and $\mu - e$ transition in nuclei get the main contribution from the $\ell_i \to \ell_j \gamma$ penguin, with the photon attached to a lepton pair or a quark pair. Such diagrams are enhanced by a $\tan\beta$ factor compared to box diagrams, and will dominate.  As a result, simple relations for their branching ratios
can be derived \cite{Paradisi:2005fk}: $Br(\mu -e \;\text{in Ti}) \approx 6\times 10^{-3} Br(\mu\to e\gamma)$ and  $Br(\ell_i \to 3\ell_j) \approx 7\times 10^{-3} Br(\ell_i \to \ell_j \gamma)$. Future experiments can probe $\mu - e$ transition in nuclei  that will be sensitive to the level of $O(10^{-18})$. These models  can be experimentally probed by future experiments in $\mu-e$ conversion in nuclei, $\mu \to 3e$ decay as well as $\mu \to e\gamma$ transition.


\section{Discussion and conclusion}

In this paper we have presented a $U(1)_{\rm PQ}$ embedding of SUSY $SO(10)$ models.  We have adopted renormalizable $SO(10)$ with a minimal Yukawa sector consisting of two matrices in flavor space.  Such a Yukawa structure has been known to fit all of fermion masses and mixings, including neutrino oscillation data, in terms of a relatively small number of parameters.  Without the PQ symmetry, these models would require a mini-split SUSY spectrum with gauginos having masses of order TeV and sfermions having masses in the 100 TeV range.  Successful implementation of the PQ symmetry enables us to lower the SUSY spectrum, which now allows all SUSY particles to have masses of order TeV.  Proton decay rates are suppressed in this scenario by a factor of $(M_{\rm PQ}/M_{\rm GUT})^2 \sim 10^{-10}$ compared to the same model without the PQ symmetry.

One feature of the PQ embedding is the appearance of an extra pair of Higgs doublets at the PQ scale.  We have shown that unification of gauge couplings can be maintained by small threshold effects arising from color sextet scalars near the GUT scale.  Although suppressed relative to SUSY $SO(10)$ models without the PQ symmetry, proton lifetime is still within observable range. Both the Higgsino mediated $d=5$ decay mode $p \rightarrow \overline{\nu} K^+$ and the gauge boson mediated $d=6$ mode $p \rightarrow e^+ \pi^0$ are within reach of ongoing and forthcoming experiments, with the rate for $e^+\pi^0$ mode enhanced due to the lowering of the GUT scale in the PQ model compared to other GUTs. 

The proposed model also predicts observable lepton flavor violation, notably in the decay $\mu \to e \gamma$ and in $\mu-e$ conversion in nuclei.  These processes originate from the Dirac neutrino Yukawa couplings which are effective in the momentum range $10^{12}$ GeV $\leq \mu \leq 10^{16}$ GeV, which transfer the LFV information to the sleptons and in turn to the leptons. The flavor structure of these LFV operators is completely determined in the model in terms of fermion masses and mixings.

\section*{Acknowledgments}
This work is supported in part by the U.S. Department of Energy Grant No.  de-sc0016013  (K.S.B). We thank Borut Bajc for useful discussions.



\begin{thebibliography}{99}


\bibitem{Pati:1974yy}
  J.~C.~Pati and A.~Salam,
  ``Lepton Number as the Fourth Color,''
  Phys.\ Rev.\ D {\bf 10}, 275 (1974)
  Erratum: [Phys.\ Rev.\ D {\bf 11}, 703 (1975)].


\bibitem{Georgi:1974sy}
  H.~Georgi and S.~L.~Glashow,
  ``Unity of All Elementary Particle Forces,''
  Phys.\ Rev.\ Lett.\  {\bf 32}, 438 (1974).


\bibitem{Georgi:1974yf}
  H.~Georgi, H.~R.~Quinn and S.~Weinberg,
  ``Hierarchy of Interactions in Unified Gauge Theories,''
  Phys.\ Rev.\ Lett.\  {\bf 33}, 451 (1974).

\bibitem{so101}

H. Georgi, in Particles and Fields, AIP, New York (1975), p. 575 (C.E. Carlson, Ed).

\bibitem{so102}

  H.~Fritzsch and P.~Minkowski,
  ``Unified Interactions of Leptons and Hadrons,''
  Annals Phys.\  {\bf 93}, 193 (1975).


\bibitem{Babu:1992ia} 
  K.~S.~Babu and R.~N.~Mohapatra,
  ``Predictive neutrino spectrum in minimal SO(10) grand unification,''
  Phys.\ Rev.\ Lett.\  {\bf 70}, 2845 (1993)
  [hep-ph/9209215].
\bibitem{Bajc:2002iw}
  B.~Bajc, G.~Senjanovic and F.~Vissani,
  ``b - tau unification and large atmospheric mixing: A Case for noncanonical seesaw,''
  Phys.\ Rev.\ Lett.\  {\bf 90}, 051802 (2003)
  [hep-ph/0210207].
\bibitem{Fukuyama:2002ch}
  T.~Fukuyama and N.~Okada,
  ``Neutrino oscillation data versus minimal supersymmetric SO(10) model,''
  JHEP {\bf 0211}, 011 (2002),
  [hep-ph/0205066].
\bibitem{Goh:2003sy}
  H.~S.~Goh, R.~N.~Mohapatra and S.~P.~Ng,
  ``Minimal SUSY SO(10), b tau unification and large neutrino mixings,''
  Phys.\ Lett.\ B {\bf 570}, 215 (2003),
  [hep-ph/0303055].
\bibitem{Goh:2003hf}
  H.~S.~Goh, R.~N.~Mohapatra and S.~P.~Ng,
  ``Minimal SUSY SO(10) model and predictions for neutrino mixings and leptonic CP violation,''
  Phys.\ Rev.\ D {\bf 68}, 115008 (2003),
  [hep-ph/0308197].
\bibitem{Bertolini:2004eq}
  S.~Bertolini, M.~Frigerio and M.~Malinsky,
  ``Fermion masses in SUSY SO(10) with type II seesaw: A Non-minimal predictive scenario,''
  Phys.\ Rev.\ D {\bf 70}, 095002 (2004),
  [hep-ph/0406117].
\bibitem{Babu:2005ia}
  K.~S.~Babu and C.~Macesanu,
  ``Neutrino masses and mixings in a minimal SO(10) model,''
  Phys.\ Rev.\ D {\bf 72}, 115003 (2005),
  [hep-ph/0505200].
\bibitem{Bertolini:2006pe}
  S.~Bertolini, T.~Schwetz and M.~Malinsky,
  ``Fermion masses and mixings in SO(10) models and the neutrino challenge to SUSY GUTs,''
  Phys.\ Rev.\ D {\bf 73}, 115012 (2006),
  [hep-ph/0605006].
\bibitem{Joshipura:2011nn}
  A.~S.~Joshipura and K.~M.~Patel,
  ``Fermion Masses in SO(10) Models,''
  Phys.\ Rev.\ D {\bf 83}, 095002 (2011),
  [arXiv:1102.5148 [hep-ph]].
\bibitem{Altarelli:2013aqa}
  G.~Altarelli and D.~Meloni,
  ``A non supersymmetric SO(10) grand unified model for all the physics below $M_{GUT}$,''
  JHEP {\bf 1308}, 021 (2013),
  [arXiv:1305.1001 [hep-ph]].
\bibitem{Dueck:2013gca}
  A.~Dueck and W.~Rodejohann,
  ``Fits to SO(10) Grand Unified Models,''
  JHEP {\bf 1309}, 024 (2013),
  [arXiv:1306.4468 [hep-ph]].
\bibitem{Bajc:2008dc}
  B.~Bajc, I.~Dorsner and M.~Nemevsek,
  ``Minimal SO(10) splits supersymmetry,''
  JHEP {\bf 0811}, 007 (2008),
  [arXiv:0809.1069 [hep-ph]].
\bibitem{Fukuyama:2015kra} 
  T.~Fukuyama, K.~Ichikawa and Y.~Mimura,
  ``Revisiting fermion mass and mixing fits in the minimal SUSY $SO(10)$ GUT,''
  Phys.\ Rev.\ D {\bf 94}, no. 7, 075018 (2016)
  [arXiv:1508.07078 [hep-ph]].
\bibitem{Fukuyama:2016vgi} 
  T.~Fukuyama, K.~Ichikawa and Y.~Mimura,
  ``Relation between proton decay and PMNS phase in the minimal SUSY $SO(10)$ GUT,''
  Phys.\ Lett.\ B {\bf 764}, 114 (2017)
  [arXiv:1609.08640 [hep-ph]].
\bibitem{Babu:2018tfi} 
  K.~S.~Babu, B.~Bajc and S.~Saad,
  ``Resurrecting Minimal Yukawa Sector of SUSY SO(10),''
  JHEP {\bf 1810}, 135 (2018)
  [arXiv:1805.10631 [hep-ph]].
\bibitem{Deppisch:2018flu} 
  T.~Deppisch, S.~Schacht and M.~Spinrath,
  ``Confronting SUSY SO(10) with updated Lattice and Neutrino Data,''
  arXiv:1811.02895 [hep-ph].
  
\bibitem{DayaBay}
  F.~P.~An {\it et al.} [Daya Bay Collaboration],
  ``Observation of electron-antineutrino disappearance at Daya Bay,''
  Phys.\ Rev.\ Lett.\  {\bf 108}, 171803 (2012)
  [arXiv:1203.1669 [hep-ex]].

\bibitem{Peccei:1977hh} 
  R.~D.~Peccei and H.~R.~Quinn,
 ``CP Conservation in the Presence of Instantons,''
  Phys.\ Rev.\ Lett.\  {\bf 38}, 1440 (1977).
  
\bibitem{Weinberg:1977ma} 
  S.~Weinberg,
  ``A New Light Boson?,''
  Phys.\ Rev.\ Lett.\  {\bf 40}, 223 (1978).
  
\bibitem{Wilczek:1977pj} 
  F.~Wilczek,
  ``Problem of Strong  $P$  and  $T$  Invariance in the Presence of Instantons,''
  Phys.\ Rev.\ Lett.\  {\bf 40}, 279 (1978).
  
\bibitem{Kim:1979if} 
  J.~E.~Kim,
  ``Weak Interaction Singlet and Strong CP Invariance,''
  Phys.\ Rev.\ Lett.\  {\bf 43}, 103 (1979).
\bibitem{Shifman:1979if} 
  M.~A.~Shifman, A.~I.~Vainshtein and V.~I.~Zakharov,
  ``Can Confinement Ensure Natural CP Invariance of Strong Interactions?,''
  Nucl.\ Phys.\ B {\bf 166}, 493 (1980).

\bibitem{Zhitnitsky:1980tq} 
  A.~R.~Zhitnitsky,
  ``On Possible Suppression of the Axion Hadron Interactions. (In Russian),''
  Sov.\ J.\ Nucl.\ Phys.\  {\bf 31}, 260 (1980)
  [Yad.\ Fiz.\  {\bf 31}, 497 (1980)].
\bibitem{Dine:1981rt} 
  M.~Dine, W.~Fischler and M.~Srednicki,
  ``A Simple Solution to the Strong CP Problem with a Harmless Axion,''
  Phys.\ Lett.\  {\bf 104B}, 199 (1981).
\bibitem{Kim:1986ax} 
  J.~E.~Kim,
  ``Light Pseudoscalars, Particle Physics and Cosmology,''
  Phys.\ Rept.\  {\bf 150}, 1 (1987).


\bibitem{bk}

  K.~S.~Babu and S.~Khan,
  ``Minimal nonsupersymmetric $SO(10)$ model: Gauge coupling unification, proton decay, and fermion masses,''
  Phys.\ Rev.\ D {\bf 92}, no. 7, 075018 (2015)
  [arXiv:1507.06712 [hep-ph]].
  
\bibitem{ringwald}
  A.~Ernst, A.~Ringwald and C.~Tamarit,
  ``Axion Predictions in $SO(10)\times U(1)_{\rm PQ}$ Models,''
  JHEP {\bf 1802}, 103 (2018)
  [arXiv:1801.04906 [hep-ph]].



\bibitem{Hisano:1992ne} 
  J.~Hisano, H.~Murayama and T.~Yanagida,
  ``Peccei-Quinn symmetry and suppression of nucleon decay rates in SUSY GUTs,''
  Phys.\ Lett.\ B {\bf 291}, 263 (1992).
  
  
  
   
\bibitem{Weinberg:1981wj} 
  S.~Weinberg,
  ``Supersymmetry at Ordinary Energies. 1. Masses and Conservation Laws,''
  Phys.\ Rev.\ D {\bf 26}, 287 (1982).
\bibitem{Sakai:1981pk} 
  N.~Sakai and T.~Yanagida,
  ``Proton Decay in a Class of Supersymmetric Grand Unified Models,''
  Nucl.\ Phys.\ B {\bf 197}, 533 (1982).
  
  
\bibitem{Bae:2015rra} 
  K.~J.~Bae, H.~Baer, A.~Lessa and H.~Serce,
  ``Mixed axion-wino dark matter,''
  Front.\ in Phys.\  {\bf 3}, 49 (2015)
  [arXiv:1502.07198 [hep-ph]].


\bibitem{Bae:2017hlp} 
  K.~J.~Bae, H.~Baer and H.~Serce,
  ``Prospects for axion detection in natural SUSY with mixed axion-higgsino dark matter: back to invisible?,''
  JCAP {\bf 1706}, no. 06, 024 (2017)
  [arXiv:1705.01134 [hep-ph]].

\bibitem{Choi:2013lwa} 
  K.~Y.~Choi, J.~E.~Kim and L.~Roszkowski,
  ``Review of axino dark matter,''
  J.\ Korean Phys.\ Soc.\  {\bf 63}, 1685 (2013)
  [arXiv:1307.3330 [astro-ph.CO]].



  
  
  
  
\bibitem{Aulakh:1982sw}
  C.~S.~Aulakh and R.~N.~Mohapatra,
  ``Implications of Supersymmetric SO(10) Grand Unification,''
  Phys.\ Rev.\ D {\bf 28}, 217 (1983),
\bibitem{Clark:1982ai}
  T.~E.~Clark, T.~K.~Kuo and N.~Nakagawa,
  ``A So(10) Supersymmetric Grand Unified Theory,''
  Phys.\ Lett.\ B {\bf 115}, 26 (1982),
  
  
\bibitem{Aulakh:2003kg}
  C.~S.~Aulakh, B.~Bajc, A.~Melfo, G.~Senjanovic and F.~Vissani,
  ``The Minimal supersymmetric grand unified theory,''
  Phys.\ Lett.\ B {\bf 588}, 196 (2004),
  [hep-ph/0306242].
  
 

  
  
\bibitem{Aulakh:2005bd} 
  C.~S.~Aulakh,
  ``MSGUTs from germ to bloom: Towards falsifiability and beyond,''
  hep-ph/0506291.
\bibitem{Bajc:2005qe} 
  B.~Bajc, A.~Melfo, G.~Senjanovic and F.~Vissani,
  ``Fermion mass relations in a supersymmetric SO(10) theory,''
  Phys.\ Lett.\ B {\bf 634}, 272 (2006)
  [hep-ph/0511352].
\bibitem{Aulakh:2005mw} 
  C.~S.~Aulakh and S.~K.~Garg,
  ``MSGUT : From bloom to doom,''
  Nucl.\ Phys.\ B {\bf 757}, 47 (2006)
  [hep-ph/0512224].

  
  
  

\bibitem{Dutta:2004zh}
B.~Dutta, Y.~Mimura and R.~N.~Mohapatra,
``Suppressing Proton Decay in the Minimal SO(10) Model,''
Phys.\ Rev.\ Lett.\ {\bf 94} (2005) 091804
[hep-ph/0412105].

\bibitem{Mohapatra:2018biy}
R.~N.~Mohapatra and M.~Severson,
``Leptonic $CP$ Violation and Proton Decay in SUSY SO(10),''
arXiv:1805.05776 [hep-ph].


\bibitem{Fukuyama:2004ps} 
  T.~Fukuyama, A.~Ilakovac, T.~Kikuchi, S.~Meljanac and N.~Okada,
  ``SO(10) group theory for the unified model building,''
  J.\ Math.\ Phys.\  {\bf 46}, 033505 (2005)
  [hep-ph/0405300].

\bibitem{Kibble:1982dd} 
  T.~W.~B.~Kibble, G.~Lazarides and Q.~Shafi,
  ``Walls Bounded by Strings,''
  Phys.\ Rev.\ D {\bf 26}, 435 (1982).

\bibitem{Chang:1983fu} 
  D.~Chang, R.~N.~Mohapatra and M.~K.~Parida,
  ``Decoupling Parity and SU(2)-R Breaking Scales: A New Approach to Left-Right Symmetric Models,''
  Phys.\ Rev.\ Lett.\  {\bf 52}, 1072 (1984).





\bibitem{tHooft:1976rip} 
  G.~'t Hooft,
  ``Symmetry Breaking Through Bell-Jackiw Anomalies,''
  Phys.\ Rev.\ Lett.\  {\bf 37}, 8 (1976).
\bibitem{Kibble:1976sj} 
  T.~W.~B.~Kibble,
  ``Topology of Cosmic Domains and Strings,''
  J.\ Phys.\ A {\bf 9}, 1387 (1976).
\bibitem{Sikivie:1982qv} 
  P.~Sikivie,
  ``Of Axions, Domain Walls and the Early Universe,''
  Phys.\ Rev.\ Lett.\  {\bf 48}, 1156 (1982).

\bibitem{Abbott:1982af} 
  L.~F.~Abbott and P.~Sikivie,
  ``A Cosmological Bound on the Invisible Axion,''
  Phys.\ Lett.\ B {\bf 120}, 133 (1983)
  [Phys.\ Lett.\  {\bf 120B}, 133 (1983)].
\bibitem{Preskill:1982cy} 
  J.~Preskill, M.~B.~Wise and F.~Wilczek,
  ``Cosmology of the Invisible Axion,''
  Phys.\ Lett.\ B {\bf 120}, 127 (1983)
  [Phys.\ Lett.\  {\bf 120B}, 127 (1983)].
\bibitem{Dine:1982ah} 
  M.~Dine and W.~Fischler,
  ``The Not So Harmless Axion,''
  Phys.\ Lett.\ B {\bf 120}, 137 (1983)
  [Phys.\ Lett.\  {\bf 120B}, 137 (1983)].
\bibitem{Davis:1985pt} 
  R.~L.~Davis,
  ``Goldstone Bosons in String Models of Galaxy Formation,''
  Phys.\ Rev.\ D {\bf 32}, 3172 (1985).
\bibitem{Davis:1986xc} 
  R.~L.~Davis,
  ``Cosmic Axions from Cosmic Strings,''
  Phys.\ Lett.\ B {\bf 180}, 225 (1986).
\bibitem{Harari:1987ht} 
  D.~Harari and P.~Sikivie,
  ``On the Evolution of Global Strings in the Early Universe,''
  Phys.\ Lett.\ B {\bf 195}, 361 (1987).
\bibitem{Vilenkin:1982ks} 
  A.~Vilenkin and A.~E.~Everett,
  ``Cosmic Strings and Domain Walls in Models with Goldstone and PseudoGoldstone Bosons,''
  Phys.\ Rev.\ Lett.\  {\bf 48}, 1867 (1982).
\bibitem{Vilenkin:1986ku} 
  A.~Vilenkin and T.~Vachaspati,
  ``Radiation of Goldstone Bosons From Cosmic Strings,''
  Phys.\ Rev.\ D {\bf 35}, 1138 (1987).
\bibitem{Pi:1984pv} 
  S.~Y.~Pi,
  ``Inflation Without Tears,''
  Phys.\ Rev.\ Lett.\  {\bf 52}, 1725 (1984).
\bibitem{Axenides:1983hj} 
  M.~Axenides, R.~H.~Brandenberger and M.~S.~Turner,
  ``Development of Axion Perturbations in an Axion Dominated Universe,''
  Phys.\ Lett.\  {\bf 126B}, 178 (1983).
\bibitem{Seckel:1985tj} 
  D.~Seckel and M.~S.~Turner,
  ``Isothermal Density Perturbations in an Axion Dominated Inflationary Universe,''
  Phys.\ Rev.\ D {\bf 32}, 3178 (1985).
\bibitem{Linde:1985yf} 
  A.~D.~Linde,
  ``Generation of Isothermal Density Perturbations in the Inflationary Universe,''
  Phys.\ Lett.\  {\bf 158B}, 375 (1985).
\bibitem{Linde:1990yj} 
  A.~D.~Linde and D.~H.~Lyth,
  ``Axionic domain wall production during inflation,''
  Phys.\ Lett.\ B {\bf 246}, 353 (1990).
\bibitem{Turner:1990uz} 
  M.~S.~Turner and F.~Wilczek,
  ``Inflationary axion cosmology,''
  Phys.\ Rev.\ Lett.\  {\bf 66}, 5 (1991).
\bibitem{Linde:1991km} 
  A.~D.~Linde,
  ``Axions in inflationary cosmology,''
  Phys.\ Lett.\ B {\bf 259}, 38 (1991).
\bibitem{Lyth:1991ub} 
  D.~H.~Lyth,
  ``Axions and inflation: Sitting in the vacuum,''
  Phys.\ Rev.\ D {\bf 45}, 3394 (1992).
\bibitem{Kawasaki:2013iha} 
  M.~Kawasaki, T.~T.~Yanagida and K.~Yoshino,
  ``Domain wall and isocurvature perturbation problems in axion models,''
  JCAP {\bf 1311}, 030 (2013)
  [arXiv:1305.5338 [hep-ph]].
\bibitem{Kawasaki:2017kkr} 
  M.~Kawasaki and E.~Sonomoto,
  ``Domain wall and isocurvature perturbation problems in a supersymmetric axion model,''
  Phys.\ Rev.\ D {\bf 97}, no. 8, 083507 (2018)
  [arXiv:1710.07269 [hep-ph]].









\bibitem{Du:2014mqa} 
  L.~Du, X.~Li and D.~X.~Zhang,
  ``Connection between proton decay suppression and seesaw mechanism in supersymmetric SO(10) models,''
  JHEP {\bf 1410}, 36 (2014)
  [arXiv:1406.2081 [hep-ph]].
  


\bibitem{Hisano:1992jj} 
  J.~Hisano, H.~Murayama and T.~Yanagida,
  Nucl.\ Phys.\ B {\bf 402}, 46 (1993)
  doi:10.1016/0550-3213(93)90636-4
  [hep-ph/9207279].
\bibitem{Ellis:1983qm} 
  J.~R.~Ellis, J.~S.~Hagelin, D.~V.~Nanopoulos and K.~Tamvakis,
  ``Observable Gravitationally Induced Baryon Decay,''
  Phys.\ Lett.\  {\bf 124B}, 484 (1983).
\bibitem{Babu:1995cw} 
  K.~S.~Babu and S.~M.~Barr,
  ``Proton decay and realistic models of quark and lepton masses,''
  Phys.\ Lett.\ B {\bf 381}, 137 (1996)
  [hep-ph/9506261].
\bibitem{Babu:1998wi} 
  K.~S.~Babu, J.~C.~Pati and F.~Wilczek,
  ``Fermion masses, neutrino oscillations, and proton decay in the light of Super-Kamiokande,''
  Nucl.\ Phys.\ B {\bf 566}, 33 (2000)
  [hep-ph/9812538].

\bibitem{Fukuyama:2004xs} 
  T.~Fukuyama, A.~Ilakovac, T.~Kikuchi, S.~Meljanac and N.~Okada,
  ``General formulation for proton decay rate in minimal supersymmetric SO(10) GUT,''
  Eur.\ Phys.\ J.\ C {\bf 42}, 191 (2005)
  [hep-ph/0401213].

\bibitem{Goh:2003nv} 
  H.~S.~Goh, R.~N.~Mohapatra, S.~Nasri and S.~P.~Ng,
  ``Proton decay in a minimal SUSY SO(10) model for neutrino mixings,''
  Phys.\ Lett.\ B {\bf 587}, 105 (2004)
  [hep-ph/0311330].






\bibitem{ss1}
P.~Minkowski,
Phys.\ Lett.\ B {\bf 67} (1977) 421;
T.~Yanagida, proceedings of the {\em Workshop on Unified Theories
and Baryon Number in the Universe}, Tsukuba, 1979, eds.
A. Sawada, A. Sugamoto.
\bibitem{ss2}
S.~Glashow, in {\em Cargese 1979, Proceedings, Quarks and Leptons}
(1979).
\bibitem{ss3}
M.~Gell-Mann, P.~Ramond, R.~Slansky, proceedings of the
{\em Supergravity Stony Brook Workshop}, New York, 1979,
eds. P. Van Niewenhuizen, D. Freeman.
\bibitem{ss4}
R.~Mohapatra, G.~Senjanovi\' c,
``Neutrino Mass and Spontaneous Parity Violation,''
Phys.Rev.Lett. {\bf 44} (1980) 912.



\bibitem{Antusch:2013jca}
  S.~Antusch and V.~Maurer,
  ``Running quark and lepton parameters at various scales,''
  JHEP {\bf 1311}, 115 (2013)
  [arXiv:1306.6879 [hep-ph]].

\bibitem{deSalas:2017kay}
  P.~F.~de Salas, D.~V.~Forero, C.~A.~Ternes, M.~Tortola and J.~W.~F.~Valle,
  ``Status of neutrino oscillations 2018: first hint for normal mass ordering and improved CP sensitivity,''
  arXiv:1708.01186 [hep-ph].

\bibitem{Machacek:1983fi}
  M.~E.~Machacek and M.~T.~Vaughn,
 ``Two Loop Renormalization Group Equations in a General Quantum Field Theory. 2. Yukawa Couplings,''
  Nucl.\ Phys.\ B {\bf 236}, 221 (1984).

\bibitem{Arason:1991ic}
  H.~Arason, D.~J.~Castano, B.~Keszthelyi, S.~Mikaelian, E.~J.~Piard, P.~Ramond and B.~D.~Wright,
  ``Renormalization group study of the standard model and its extensions. 1. The Standard model,''
  Phys.\ Rev.\ D {\bf 46}, 3945 (1992).

\bibitem{Babu:1987im}
  K.~S.~Babu,
  ``Renormalization Group Analysis of the {Kobayashi-Maskawa} Matrix,''
  Z.\ Phys.\ C {\bf 35}, 69 (1987).

\bibitem{Babu:1993qv}
  K.~S.~Babu, C.~N.~Leung and J.~T.~Pantaleone,
  ``Renormalization of the neutrino mass operator,''
  Phys.\ Lett.\ B {\bf 319}, 191 (1993)
  [hep-ph/9309223].

\bibitem{Chankowski:1993tx} 
  P.~H.~Chankowski and Z.~Pluciennik,
  ``Renormalization group equations for seesaw neutrino masses,''
  Phys.\ Lett.\ B {\bf 316}, 312 (1993)
  [hep-ph/9306333].

\bibitem{Antusch:2001ck}
  S.~Antusch, M.~Drees, J.~Kersten, M.~Lindner and M.~Ratz,
  ``Neutrino mass operator renormalization revisited,''
  Phys.\ Lett.\ B {\bf 519}, 238 (2001)
  [hep-ph/0108005].

\bibitem{Barger:1992ac}
  V.~D.~Barger, M.~S.~Berger and P.~Ohmann,
  ``Supersymmetric grand unified theories: Two loop evolution of gauge and Yukawa couplings,''
  Phys.\ Rev.\ D {\bf 47}, 1093 (1993)
  [hep-ph/9209232].

\bibitem{Barger:1992pk}
  V.~D.~Barger, M.~S.~Berger and P.~Ohmann,
  ``Universal evolution of CKM matrix elements,''
  Phys.\ Rev.\ D {\bf 47}, 2038 (1993)
  [hep-ph/9210260].

\bibitem{Antusch:2001vn}
  S.~Antusch, M.~Drees, J.~Kersten, M.~Lindner and M.~Ratz,
  ``Neutrino mass operator renormalization in two Higgs doublet models and the MSSM,''
  Phys.\ Lett.\ B {\bf 525}, 130 (2002)
  [hep-ph/0110366].


\bibitem{Fukuyama:2016mqb} 
  T.~Fukuyama, N.~Okada and H.~M.~Tran,
  ``Sparticle spectroscopy of the minimal SO(10) model,''
  Phys.\ Lett.\ B {\bf 767}, 295 (2017)
  [arXiv:1611.08341 [hep-ph]].

\bibitem{kamlandzen}

  A.~Gando {\it et al.} [KamLAND-Zen Collaboration],
  ``Search for Majorana Neutrinos near the Inverted Mass Hierarchy Region with KamLAND-Zen,''
  Phys.\ Rev.\ Lett.\  {\bf 117}, no. 8, 082503 (2016)
  Addendum: [Phys.\ Rev.\ Lett.\  {\bf 117}, no. 10, 109903 (2016)]
  [arXiv:1605.02889 [hep-ex]].

\bibitem{Babu:2016cri} 
  K.~S.~Babu, B.~Bajc and S.~Saad,
  ``New Class of SO(10) Models for Flavor,''
  Phys.\ Rev.\ D {\bf 94}, no. 1, 015030 (2016)
  [arXiv:1605.05116 [hep-ph]].


\bibitem{Hayato:1999az} 
  V.~Takhistov [Super-Kamiokande Collaboration],
  ``Review of Nucleon Decay Searches at Super-Kamiokande,''
  arXiv:1605.03235 [hep-ex].

\bibitem{Aoki:2017puj} 
  Y.~Aoki, T.~Izubuchi, E.~Shintani and A.~Soni,
  ``Improved lattice computation of proton decay matrix elements,''
  Phys.\ Rev.\ D {\bf 96}, no. 1, 014506 (2017)
  [arXiv:1705.01338 [hep-lat]].



\bibitem{Babu:2010ej} 
  K.~S.~Babu, J.~C.~Pati and Z.~Tavartkiladze,
  ``Constraining Proton Lifetime in SO(10) with Stabilized Doublet-Triplet Splitting,''
  JHEP {\bf 1006}, 084 (2010)
  [arXiv:1003.2625 [hep-ph]].

\bibitem{sk2}
  K.~Abe {\it et al.} [Super-Kamiokande Collaboration],
    ``Search for proton decay cia $p \rightarrow e^+ \pi^0$ and $p \rightarrow \mu^+ \pi^0$ in 0.31 megaton exposure of Super-Kamiokande water Cherenkov detector", 
  Phys.\ Rev.\ D {\bf 95}, no. 1, 012004 (2017)
  [arXiv:1610.03597 [hep-ex]].


\bibitem{Babu:2008ge} 
  K.~S.~Babu, I.~Gogoladze, M.~U.~Rehman and Q.~Shafi,
  ``Higgs Boson Mass, Sparticle Spectrum and Little Hierarchy Problem in Extended MSSM,''
  Phys.\ Rev.\ D {\bf 78}, 055017 (2008)
  [arXiv:0807.3055 [hep-ph]].
  
\bibitem{Hisano:1995cp} 
  J.~Hisano, T.~Moroi, K.~Tobe and M.~Yamaguchi,
  ``Lepton flavor violation via right-handed neutrino Yukawa couplings in supersymmetric standard model,''
  Phys.\ Rev.\ D {\bf 53}, 2442 (1996)
  [hep-ph/9510309].
\bibitem{Paradisi:2005fk} 
  P.~Paradisi,
  ``Constraints on SUSY lepton flavor violation by rare processes,''
  JHEP {\bf 0510}, 006 (2005)
  [hep-ph/0505046].
\bibitem{Ciuchini:2007ha} 
  M.~Ciuchini, A.~Masiero, P.~Paradisi, L.~Silvestrini, S.~K.~Vempati and O.~Vives,
  ``Soft SUSY breaking grand unification: Leptons versus quarks on the flavor playground,''
  Nucl.\ Phys.\ B {\bf 783}, 112 (2007)
  [hep-ph/0702144 [HEP-PH]].
\bibitem{TheMEG:2016wtm} 
  A.~M.~Baldini {\it et al.} [MEG Collaboration],
  ``Search for the lepton flavour violating decay $\mu ^+ \rightarrow \mathrm {e}^+ \gamma $ with the full dataset of the MEG experiment,''
  Eur.\ Phys.\ J.\ C {\bf 76}, no. 8, 434 (2016)
  [arXiv:1605.05081 [hep-ex]].







\end{thebibliography}
\end{document}